\documentclass[a4paper,11pt]{article}
\pdfoutput=1 

\usepackage{jcappub} 

\usepackage[T1]{fontenc} 
\usepackage{caption}
\usepackage{subcaption}

\usepackage[latin9]{inputenc}
\usepackage{float}
\usepackage{amsmath}
\usepackage{amssymb}
\usepackage{graphicx}
\usepackage{esint}
 \usepackage{hyperref}
\usepackage{comment}
\usepackage{color}
\usepackage{microtype}
\usepackage{cleveref}
\usepackage{breakurl}
\usepackage{bbm}
\newcommand{\be}{\begin{equation}}
\newcommand{\ee}{\end{equation}}
\newcommand{\ben}{\begin{displaymath}}
\newcommand{\een}{\end{displaymath}}
\newcommand{\bea}{\begin{eqnarray}}
\newcommand{\eea}{\end{eqnarray}}
\def\K{K{\"a}hler}
   \newcommand{\rf}[1]{(\ref{#1})}

\def\be{\begin{equation}}
\def\ee{\end{equation}}
\def\bea{\begin{eqnarray}}
\def\eea{\end{eqnarray}}
\def\ba{\begin{array}}
\def\ea{\end{array}}
\def\bit{\begin{itemize}}
\def\eit{\end{itemize}}

\def\la{\lambda}

\newcommand{\cN}{\mathcal{N}}

 \makeatletter

\@ifundefined{textcolor}{}{%
 \definecolor{BLACK}{gray}{0}
 \definecolor{WHITE}{gray}{1}
 \definecolor{RED}{rgb}{1,0,0}
 \definecolor{GREEN}{rgb}{0,1,0}
 \definecolor{BLUE}{rgb}{0,0,1}
 \definecolor{CYAN}{cmyk}{1,0,0,0}
 \definecolor{MAGENTA}{cmyk}{0,1,0,0}
 \definecolor{YELLOW}{cmyk}{0,0,1,0}
}

\newcommand{\dd}{\mathrm{d}}

\allowdisplaybreaks

\makeatother

\makeatletter
\DeclareRobustCommand{\rcite}[1]{%
  \rcite@aux#1,\@nil{#1}%
}
\def\rcite@aux#1,#2\@nil#3{%
  \if\relax#2\relax
    Ref.~\cite{#3}%
  \else
    Refs.~\cite{#3}%
  \fi
}
\makeatother

\hypersetup{
    colorlinks = true,
    citecolor = {blue},
    linkcolor = {blue},
    urlcolor = {blue},
}

 \title{ \boldmath{\rm \bf \hskip 1.5 cm The landscape, the swampland \\
 \hskip 1 cm and  the era of precision cosmology}}

\author[a]{\rm  Yashar Akrami,}
\author[b]{\rm  Renata Kallosh,}
\author[b]{\rm Andrei Linde}
\author[a, c]{\rm  and Valeri Vardanyan}

\affiliation[a]{Lorentz Institute for Theoretical Physics, Leiden University, P.O. Box 9506, 2300 RA Leiden, The Netherlands}
\affiliation[b]{Stanford Institute for Theoretical Physics and Department of Physics, Stanford University, Stanford, CA 94305, USA}
\affiliation[c]{Leiden Observatory, Leiden University, P.O. Box 9513, 2300 RA Leiden, The Netherlands}

\emailAdd{akrami@lorentz.leidenuniv.nl}
\emailAdd{kallosh@stanford.edu}
\emailAdd{alinde@stanford.edu}
\emailAdd{vardanyan@lorentz.leidenuniv.nl}

\abstract{ 
We review the advanced version of the KKLT construction and pure $d=4$ de~Sitter supergravity, involving  a nilpotent multiplet,  with regard  to  various  conjectures that de Sitter state cannot exist in string theory. We  explain why we consider these conjectures problematic and not well motivated, and why the  recently proposed  alternative string theory models of dark energy, ignoring vacuum stabilization, are ruled out by cosmological observations at least at the $3\sigma$ level, i.e. with more than $99.7\%$ confidence.}


\begin{document}
 \maketitle


 
\section{Introduction}\label{sec:intro}
\parskip 6pt

The observation of late-time cosmic acceleration, almost exactly 20 years ago, is one of the most important cosmological discoveries of all time.   As a result of that, we now face two extremely difficult problems at once: we have to explain why the vacuum energy/cosmological constant $\Lambda$ is not exactly zero but is extremely small, about $0.7\times 10^{{-120}}$ in $d=4$ Planck units, and why it is of the same order  as the density of normal matter in the universe, but only at the present epoch. This problem was addressed by constructing $d=4$ de Sitter (dS) vacua in the context of KKLT construction in Type IIB superstring theory~\cite{Kachru:2003aw,Kachru:2003sx}. De Sitter vacua in noncritical string theory were studied earlier in~\cite{Silverstein:2001xn,Maloney:2002rr}.

\begin{figure}[t!]
 \begin{center}
\includegraphics[scale=0.34]{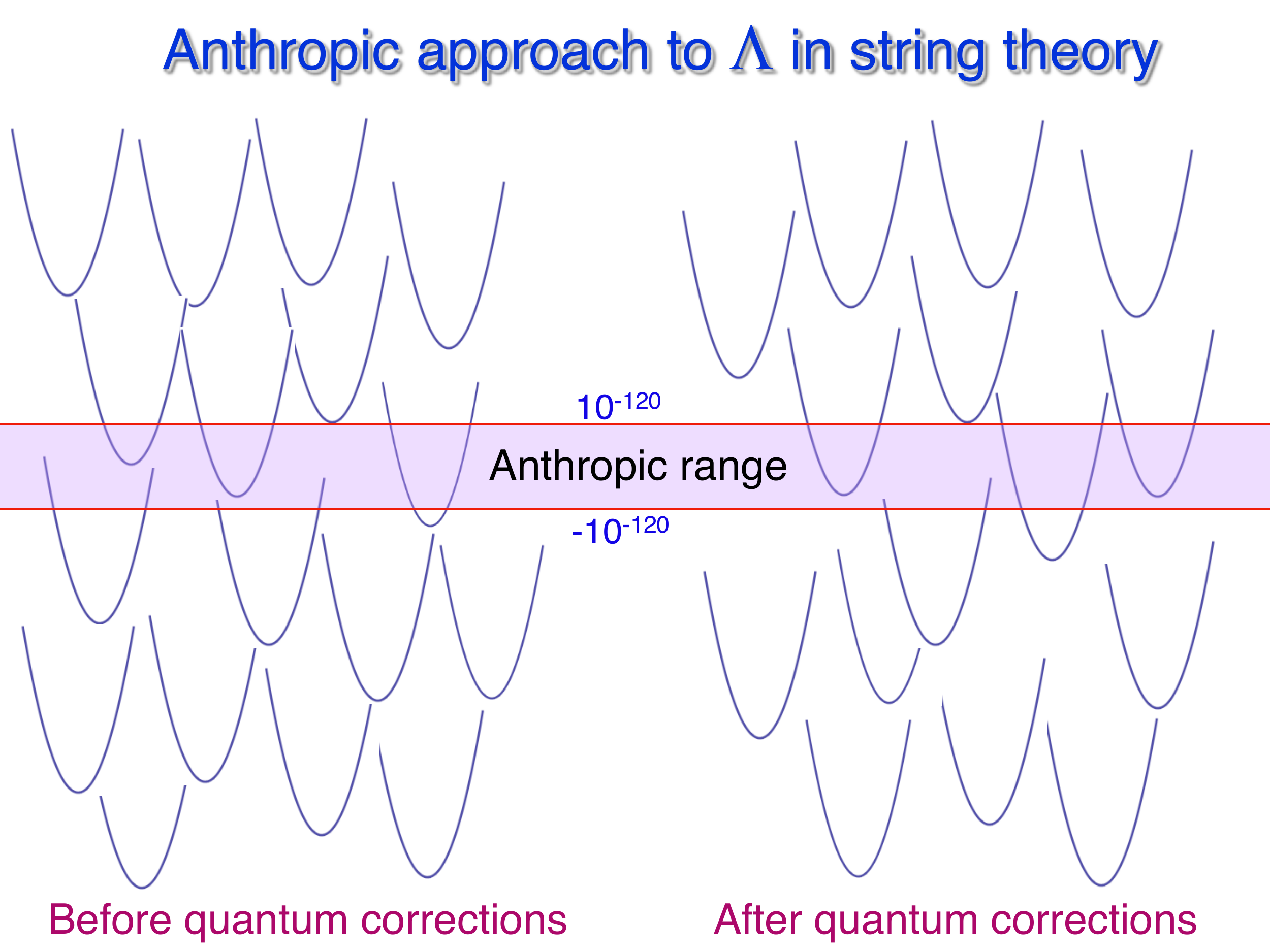}
 \end{center}
\hskip 1cm \vskip -0.7cm
\caption{\footnotesize There are many vacua before quantum corrections, and many vacua after quantum corrections. The ones on the right in the anthropically allowed range may originate from the ones on the left which were at all possible values of $\Lambda$. In this picture, quantum corrections may be large or small, but there will still be some vacua in the  anthropic  range,  after all possible quantum corrections are made.}
\label{chi}
\end{figure}

The most important part of de Sitter constructions in string theory and its various generalizations is the enormous combinatorial multiplicity of vacuum states in the theory~\cite{Douglas:2003um,Douglas:2006es,Denef:2007pq} and the possibility to tunnel from one of these states to another in the string theory landscape~\cite{Kachru:2003aw,Susskind:2003kw}, just as anticipated in the eternal chaotic inflation scenario~\cite{Linde:1986fd}. 
The value of the cosmological constant $\Lambda$ originates from an incomplete cancellation between two contributions to energy, a negative one, $V_{\rm AdS}< 0,$ due to the Anti-de Sitter (AdS) minimum used for moduli stabilization, and a positive one, due to an $\overline {D3}$-brane. In different parts (or different quantum states) of the universe, the difference between these two may take arbitrary values, but in the part of the universe where we can live it must be extremely small~\cite{Davies:1980ji,Linde:1984ir,Sakharov:1984ir,Banks:1984cw,Linde:1986fd,Barrow:1988yia,Linde:1986dq,Weinberg:1987dv,Martel:1997vi,Garriga:1999hu,Bousso:2000xa,Susskind:2003kw,Linde:2015edk}, 
\be
\Lambda = V_{\overline {D3}}-  V_{\rm AdS} \approx 10^{{-120}} \ .
\label{dif}\ee
Quantum corrections may affect  vacuum energy in each of the dS or AdS minima, but one may argue that if the total number of possible vacua is large enough, there will be many vacua where the cosmological constant belongs to the anthropically allowed range $|\Lambda | \lesssim 10^{{-120}}$, as we have depicted in  Fig. \ref{chi}. This makes the anthropic solution of the cosmological constant problem in the context of the string theory landscape rather robust. 

Although the basic features of the string landscape theory were formulated long ago, the progress in this direction still continues. Many interesting generalizations of the KKLT scenario have been proposed, some of which are mentioned below. Simultaneously, there have been many attempts to disprove the concept of string theory landscape, to prove that de Sitter vacua in string theory cannot be stable or metastable, and to provide an alternative solution to the cosmological constant problem.  However, despite a significant effort during the last 15 years, no compelling alternative solution to the cosmological constant problem has been found as yet. 

Recently, a new attempt has been made in~\cite{Obied:2018sgi}. The authors conjectured that stable or metastable de Sitter vacua could not exist in string theory, and suggested to return to the development of superstring theory versions of quintessence models, simultaneously imposing a strong (and, in our opinion, not well motivated) 
constraint  on quintessence models, ${| \nabla_\phi V|\over V}  \geq c \sim 1$. The list of the currently available models of this type is given in~\cite{Obied:2018sgi}, and their cosmological consequences are studied in~\cite{Agrawal:2018own}, where a confusing conclusion has been drawn.

In the abstract of~\cite{Agrawal:2018own} one finds: ``We study constraints imposed by two proposed string Swampland criteria on cosmology \dots~Applying these same criteria to dark energy in the present epoch, we find that specific quintessence models can satisfy these bounds and, at the same time, satisfy current observational constraints.''
However, in  Section 5 of the same paper one reads: ``notably there do not exist rigorously proven examples in hand where $c$ is as small as $0.6$, as required to satisfy current observational constraints on dark energy.''  Indeed,~\cite{Agrawal:2018own} has argued that the models with $c > 0.6$ are ruled out at the $3\sigma$ level, even though in the revised version of the paper they say that  $c > 0.6$ is ruled out at the $2\sigma$ level.  An additional uncertainty has been introduced by Heisenberg et al.~\cite{Heisenberg:2018yae}, who claim that models with $c \leq 1.35$ are consistent with observations. 

The first goal of the present paper is to explain that the `no-dS'  conjecture of ~\cite{Brennan:2017rbf,Danielsson:2018ztv,Obied:2018sgi} is based, in part, on the  no-go theorem~\cite{Maldacena:2000mw}, which has already been addressed in the KKLT construction~\cite{Kachru:2003aw,Kachru:2003sx}. Important  developments in the KKLT construction during the last 4 years \cite{Kallosh:2014wsa,Ferrara:2014kva,Bergshoeff:2015jxa,Bergshoeff:2015tra,Hasegawa:2015bza,Kallosh:2015nia,Kallosh:2016aep,Aalsma:2018pll}, including the theory of uplifting from AdS to dS and the discovery of dS supergravity~\cite{Bergshoeff:2015tra,Hasegawa:2015bza} which addressed another no-go theorem~\cite{Pilch:1984aw}, 
are not even mentioned in \cite{Brennan:2017rbf,Obied:2018sgi,Agrawal:2018own}, as well as in a recent review on compactification in string theory \cite{Danielsson:2018ztv}.

The second goal of our paper is to determine which of the three 
conclusions of  \cite{Agrawal:2018own,Heisenberg:2018yae} on the observational constraints on $c$ is correct. We show that dark energy models with $c > 1$ are ruled out at the $3\sigma$ level, i.e. with $99.7\%$ confidence. All the models discussed in \cite{Obied:2018sgi,Agrawal:2018own}, which may be qualified as derived from string theory in application to the four-dimensional (4d) universe, require $c \geq \sqrt{2} \sim 1.4$, which is  ruled out by  cosmological observations.  
If one attempts to extend the conjecture  ${| \nabla_\phi V|\over V}  \geq 1$ to inflationary models (which would be even less motivated, as discussed in Section \ref{SW}), this conjecture would be in an even stronger contradiction with the cosmological observations. 

Note that the class of string theory models studied in~\cite{Obied:2018sgi,Agrawal:2018own} includes neither non-perturbative effects, nor the effects related to the KKLT uplifting due to a single $\overline {D3}$-brane, which is described in $d=4$ supergravity by a nilpotent multiplet~\cite{Kallosh:2014wsa,Ferrara:2014kva,Bergshoeff:2015jxa,Bergshoeff:2015tra,Hasegawa:2015bza,Kallosh:2015nia,Kallosh:2016aep,Aalsma:2018pll}. Therefore, the KKLT model, as well as  available inflationary models based on string theory, involves elements which do not belong to the class of models studied in  \cite{Obied:2018sgi,Agrawal:2018own}. It is therefore not very surprising that all the models of accelerated expansion of the universe studied in~\cite{Obied:2018sgi,Agrawal:2018own} are ruled out by observational data on dark energy and inflation.

In Section~\ref{KKLT}, we briefly describe the recent progress in the KKLT construction and dS supergravity. We describe the KKLT scenario in the theory with a nilpotent multiplet, and its generalizations with strong vacuum stabilization which are especially suitable for cosmological applications. In Section~\ref{nogo}, we discuss various no-go theorems which were supposed to support the no-dS conjecture of~\cite{Brennan:2017rbf,Danielsson:2018ztv,Obied:2018sgi}, and reply to the criticism of the KKLT construction in these and other papers.  Section~\ref{fullfledged} describes the recent progress towards full-fledged string theory solutions describing dS vacua. Various versions of the no-dS conjecture are described in Section~\ref{SW}. Cosmological constraints on the parameters of quintessence models relevant to the discussions of this paper are obtained in Section~\ref{observations}, where the focus is on models with single-exponential potentials. In Section~\ref{models}, we present a detailed analysis of the string theory based quintessence models proposed in~\cite{Obied:2018sgi}, which, on the one hand, can qualify as derived from string theory compactified to $d=4$, and on the other hand, are used to support the dark energy swampland conjecture $V_{,\phi}/V\geq 1$. This does not include models of quintessence with $d\neq 4$, as well as models for which $V_{,\phi}=0$ is possible. In Section~\ref{problems}, we discuss general conceptual problems with models of quintessence in string theory.

Appendix \ref{dSsupergravity} contains a more technical discussion of no-go theorems and of the advanced KKLT construction and dS supergravity.  In Appendix \ref{supercritical}, we review quintessence models in supercritical string theory with the total number of dimensions $D\gg 26$~\cite{Dodelson:2013iba}. We compare in Appendix~\ref{sec:priorbounds} the observational bounds on the string theory models of quintessence discussed in  Section~\ref{observations} with those provided in  \cite{Agrawal:2018own,Heisenberg:2018yae}. 
Appendix~\ref{sec:doubleexp} and Appendix~\ref{sec:two-field} present a discussion of and observational constraints on double-exponential quintessence potentials, which appear in some of the string theory models of Section~\ref{models}. Finally, in Appendix \ref{pres} we give some examples illustrating the rapidly improving precision of measurements of the cosmological parameters during the last decade. It  shows that even a small difference in  some of the experimental results can make a huge difference for the development of theoretical cosmology. This is very different from the situation two decades ago, when `order-of-magnitude' theoretical predictions could be good enough.

 \section{KKLT and the string theory landscape}\label{KKLT}

The KKLT construction and its various subsequent generalizations consist of two parts. First of all, one may use  non-perturbative effects to stabilize string theory moduli, including the volume modulus responsible for compactification, in a supersymmetric AdS vacuum state. This can be done by several different methods; see, e.g.,~\cite{Kachru:2003aw,Kallosh:2004yh,Balasubramanian:2005zx}.
The second step involves uplifting of the AdS vacuum with  negative  energy density  to dS by adding a contribution of a single $\overline {D3}$-brane.  A detailed explanation of this procedure, starting with the $\overline {D3}$-brane action, has been given in~\cite{Kachru:2003sx} and corresponds to eq.~\rf{dif}. More recent string theory constructions of dS vacua, developing various corners of the string theory landscape, have been presented in~\cite{Cicoli:2013cha,Kallosh:2014oja,Cicoli:2015ylx,Gallego:2017dvd}. 
       
An advanced version of the KKLT uplifting has been developed more recently in~\cite{Kallosh:2014wsa,Ferrara:2014kva,Bergshoeff:2015jxa,Bergshoeff:2015tra,Hasegawa:2015bza,Kallosh:2015nia,Kallosh:2016aep,Aalsma:2018pll}.   It is based on a manifest, nonlinearly realized, spontaneously broken, Volkov-Akulov type supersymmetry, discovered in 1972~\cite{Volkov:1972jx}. It was observed by John Schwarz et al.~\cite{Aganagic:1996nn} back in 1997 that the local $\kappa$-symmetry of a single $\overline {D3}$ upon gauge-fixing becomes Volkov-Akulov supersymmetry.\footnote{Note that the nonlinearly realized supersymmetry on D-branes discovered in~\cite{Aganagic:1996nn} differs from the linear one. Therefore, an important prediction of non-perturbative string theory is nonlinear supersymmetry. It is supported by observational cosmology where de Sitter and near de Sitter spaces play a fundamental role.} The goldstino multiplet with a nonlinearly realized supersymmetry has only a fermion, there is no scalar partner.

The core of the uplifting from an AdS vacuum with a negative cosmological constant to a dS vacuum with a positive cosmological constant is due to the tension of a single $\overline {D3}$ represented at the level of the effective field theory by a positive energy of the goldstino action.
The string theory interpretation of vacuum stabilization and uplifting, which is an important part of the advanced KKLT construction, was supported and explained in the SUSY 2015 talk by Polchinski~\cite{Polchinski:2015bea}. In the advanced version of the KKLT construction~\cite{Kallosh:2014wsa,Ferrara:2014kva,Bergshoeff:2015jxa,Bergshoeff:2015tra,Hasegawa:2015bza,Kallosh:2015nia,Kallosh:2016aep,Aalsma:2018pll}, the single $\overline {D3}$-brane has been represented at the phenomenological supergravity level by a nilpotent goldstino multiplet; see, for example,~\cite{Cribiori:2017ngp} and references therein.
      
The uplifting procedure described in the earlier version of the KKLT construction~\cite{Kachru:2003aw,Kachru:2003sx} corresponds to an approximation of~\cite{Kallosh:2014wsa,Bergshoeff:2015jxa} where the fermionic goldstino is absent. At the supergravity level, the absence of goldstino in~\cite{Kachru:2003aw,Kallosh:2004yh} corresponds to a choice of the local supersymmetry gauge where goldstino vanishes~\cite{Kallosh:2015sea}. These facts became clear only after $d=4$ pure de Sitter supergravity, promoting the global Volkov-Akulov symmetry to the level of a local supersymmetry,  was  constructed  in 2015 in~\cite{Bergshoeff:2015tra,Hasegawa:2015bza}.

The simplest $d=4$ version of the KKLT construction~\cite{Kallosh:2014wsa,Ferrara:2014kva,Bergshoeff:2015jxa,Bergshoeff:2015tra,Hasegawa:2015bza,Kallosh:2015nia,Kallosh:2016aep,Aalsma:2018pll} is  described by the \K\ potential and superpotential
\be\label{KKLT2003}
 K = -3\log\left(T+\bar T\right)+S \bar S \, , \qquad W_\text {KKLT}= W_0 + A e^{-a T} + \mu^{2}S\,,
\ee
where $T$ is the volume  modulus and $S$ is a nilpotent chiral superfield (i.e. $S^2=0$). One may also use the ``warped'' version of the \K\ potential $K= -3\log\left(T+\bar T -S \bar S\right)$. At $\mu=0$, the potential has an AdS minimum. By increasing the parameter $\mu^2$, one can uplift this minimum to dS.

A year after the  invention of the KKLT model~\cite{Kachru:2003aw}, it was recognized that combining this model with inflation would effectively lead to an additional contribution to $\mu^{2}$, which could destabilize the volume modulus in the very early universe~\cite{Kallosh:2004yh}. The destabilization may occur at a large Hubble constant because the height of the barrier in the KKLT scenario is proportional to the square of $W_{0}$ related to the gravitino mass and the strength of supersymmetry breaking, which was often considered small. This problem disappears if supersymmetry breaking in this theory is sufficiently high. 

There are several other ways to stabilize the KKLT potential. The simplest one, proposed in~\cite{Kallosh:2004yh}, is to change the superpotential to the racetrack potential with two exponents,
\be
W_\text{KL}(T,S)  =W_{0}-Ae^{-aT}- Be^{-bT}  +   \mu^{2} S\ ,
\label{adssup}
\ee
where
\be\label{w0stab}
W_0=  -A \left({a\,A\over
b\,B}\right)^{a\over b-a} +B \left ({a\,A\over b\,B}\right) ^{b\over b-a}  \ .
\ee
For $\mu = 0$, the potential $V(T)$ has a stable supersymmetric Minkowski minimum. Adding a small correction to $W_{0}$ makes this minimum AdS. For $\mu \neq 0$, this minimum  can be easily uplifted to dS while remaining strongly stabilized \cite{Kallosh:2004yh,BlancoPillado:2005fn,Kallosh:2014oja}. Importantly, the height of the barrier in this scenario is not related to supersymmetry breaking and can be arbitrarily high. Therefore, this version of the KKLT potential, sometimes called the KL model, is especially suitable for being a part of the inflationary theory~\cite{Kallosh:2011qk,Dudas:2012wi}.


The basic idea of finding a stable supersymmetric (or near-supersymmetric) vacuum state and then uplifting it without affecting its stability can be generalized for the string theory motivated theories with many moduli. A particular example is the STU model with 
\bea\label{eq:K}
K &=& -\log (S + \bar S) -3 \log (T + \bar T)-3 \log (U + \bar U)+X \bar X \,, \nonumber \\
W &=& W_{0}+A\,(S - S_{0}) (1 - c\, e^{-a\,T}) + B\,(U - U_{0})^2 +  \mu^2 X \ ,
\eea
where $X$ is a nilpotent multiplet. For $W_{0}= \mu = 0$, the potential has a supersymmetric Minkowski minimum at $S = S_{0}$, $U = U_{0}$ and $T = {\log c\over a}$ \cite{Kallosh:2014oja}. It can be easily converted to an AdS minimum by taking a tiny constant  $W_{0}$, or uplifted to dS by taking $\mu \neq 0$. Since the required value of uplift can be extremely small, one can have a theory with a controllable level of supersymmetry breaking and strong moduli stabilization.

Yet another example is an STU model with a superpotential 
\be\label{STU2}
W = W_{\rm KL}(T,X)+ P\,(S - S_{0})^{2} + Q\,(U - U_{0})^2 \ ,
\ee
where $X$ is a nilpotent multiplet, and $P$ and $Q$ are some constants. It has a supersymmetric Minkowski vacuum with all moduli stabilized at $S = S_{0}$, $U = U_{0}$ and $T =  {1\over a-b}\ln \left ({a\,A\over b\,B}\right)$, which can be downshifted to AdS or uplifted to a strongly stabilized dS vacuum, as in the previous case \cite{Kallosh:2014oja}. Importantly, none of these potentials is destabilized during uplifting.

Thus, we have a family of well-motivated models describing many scalar fields with strongly stabilized string theory dS vacua.

\section{No-go theorems for de Sitter?}\label{nogo}

Over the last 15 years, there have been many attempts to find another mechanism of vacuum stabilization in string theory,  or to find an alternative, better way of addressing the cosmological constant problem. Most of these developments concentrated on finding other mechanisms of compactification~\cite{Kallosh:2004yh,Balasubramanian:2005zx}, or developing a simpler mechanism of uplifting \cite{Kallosh:2014wsa,Ferrara:2014kva,Bergshoeff:2015tra,Hasegawa:2015bza,Bergshoeff:2015jxa,Kallosh:2015nia,Kallosh:2016aep,Aalsma:2018pll}, but none of the efforts challenged the basic principles of the string landscape scenario. 

Another  trend was to try to find problems with this construction, and then start everything anew. But starting everything anew is not an easy task. There seems to be no unique opinion about what are `controllable' string theory models and in which duality corner of string theory one should look for phenomenological models explaining the data. Many of the statements made in string theory  are  based on the non-perturbative theory and various conjectures about  quantum gravity.
 
In this situation, one may try to rely on well-established no-go theorems, which may create a certain mindset about what is possible and what is impossible, or may point out a way towards a breakthrough. For example, long time ago there was a no-go theorem by Coleman and Mandula \cite{Coleman:1967ad} stating that space-time and internal symmetries could not be combined in any but a trivial way. This powerful no-go theorem was evaded with the discovery of supersymmetry, supergravity and string theory. 
 
Similarly, there is a Maldacena-Nunez no-go theorem \cite{Maldacena:2000mw}, which does not allow a stable dS compactification of $d=10$ supergravity under certain conditions, including a requirement of a nonsingular compactification manifold.
 A discussion of this theorem and its various generalizations can be found, e.g., in~\cite{Brennan:2017rbf,Danielsson:2018ztv,Obied:2018sgi}. The theorem was discussed on the first page  of the KKLT paper \cite{Kachru:2003aw}, where it was explained, with a reference to \cite{Giddings:2001yu}, that the KKLT construction contained novel ingredients invalidating the no-go theorem.  For example, it is well known~\cite{Acharya:2001gy} that M-theory compactification on a manifold of $G_2$ holonomy can give chiral fermions in four dimensions only if the compactification manifold  is singular. Thus, this theorem can hardly be used as a general argument against dS vacua in string theory. This would be in parallel with requiring the absence of chiral fermions, in contradiction to the Standard Model.

But this is not the only relevant no-go theorem. It is known for 33 years that the no-go theorem~\cite{Pilch:1984aw} prohibits pure  supergravity with de Sitter vacua in a theory with linearly realized supersymmetry.  For quite a while, this was considered to be a real obstacle on the way towards finding dS vacua in supergravity. However, this theorem applies only to pure supergravity without matter multiplets. It is very easy to construct dS vacua in realistic supergravity models containing scalar fields. 

The new construction of local supergravity with dS vacua~\cite{Bergshoeff:2015tra,Hasegawa:2015bza} has demonstrated that one can evade the no-go theorem~\cite{Pilch:1984aw} and construct dS vacua even  in pure supergravity without scalar fields by  including a nonlinearly realized  supersymmetry. This result, closely related to the development of the advanced versions of the KKLT construction in~\cite{Kallosh:2014wsa,Ferrara:2014kva,Bergshoeff:2015tra,Hasegawa:2015bza,Bergshoeff:2015jxa,Kallosh:2015nia,Kallosh:2016aep,Aalsma:2018pll}, was obtained 3 years ago. 
 
Meanwhile, the latest critical discussions of string theory dS construction in~\cite{Brennan:2017rbf,Danielsson:2018ztv} rely on the 33-year-old no-go theorem \cite{Pilch:1984aw}, and do not even mention the advanced versions of the KKLT construction  \cite{Kallosh:2014wsa,Ferrara:2014kva,Bergshoeff:2015tra,Hasegawa:2015bza,Bergshoeff:2015jxa,Kallosh:2015nia,Kallosh:2016aep,Aalsma:2018pll}. They miss the recent discovery that one can evade the no-go theorem~\cite{Pilch:1984aw} by introducing a single $\overline {D3}$-brane. The effect of the $\overline {D3}$-brane leads to $d=4$ dS supergravity with the nilpotent multiplet $S^2(x, \theta)=0$. 
Because of the importance of these results, we briefly review them in Appendix \ref{dSsupergravity}.

Yet another dS-related no-go theorem is discussed in the last section of~\cite{Obied:2018sgi}. It generalizes the well-known result by Farhi and Guth~\cite{Farhi:1986ty} on the impossibility of creating dS universes in a laboratory at the classical level. We do not discuss this no-go theorem in our paper since it does not  apply to the standard cosmological scenario. From our perspective, the very fact that this no-go theorem has been described in the concluding section of~\cite{Obied:2018sgi}  tells a lot about the strength of the arguments against dS vacua in string theory.

Many critical comments on the KKLT mechanism made in the papers reviewed in~\cite{Danielsson:2018ztv} are based on the studies of back reaction within the classical $d=10$ supergravity approach. However, to study the back reaction  using supergravity requires  a very large  number $p$ of $\overline {D3}$-branes, $p \gg g_{s}^{{-1}}\gg 1$. As emphasized in \cite{Michel:2014lva,Polchinski:2015bea,Kallosh:2015nia}, this approach is not valid  for the most important case of $p= 1$, i.e. for a single $\overline {D3}$-brane invariant under  local fermionic $\kappa$-symmetry, which is an essential part of the advanced  KKLT construction \cite{Kallosh:2014wsa,Ferrara:2014kva,Bergshoeff:2015jxa,Bergshoeff:2015tra,Hasegawa:2015bza,Kallosh:2015nia,Kallosh:2016aep,Aalsma:2018pll}. 

Two other recent publications have been used in~\cite{Brennan:2017rbf,Danielsson:2018ztv,Obied:2018sgi} for the justification of the no-dS conjecture. The first one  \cite{Sethi:2017phn} is discussed in the recent paper by Kachru and Trivedi \cite{Kachru:2018aqn}; we agree with their conclusions.  

The second  paper is~\cite{Moritz:2017xto}. The authors proposed a modified 4d version of the KKLT model \rf{KKLT2003}.  On the basis of this modified theory, they concluded that one cannot  uplift the AdS vacuum to dS in their version of the KKLT scenario. Simultaneously, they developed a 10d version of the KKLT construction, and came to the same conclusion. However, the calculations in the 10d version were based on the results obtained by other authors in a very different context, as well as on several unproved conjectures. Therefore, the 10d analysis of ~\cite{Moritz:2017xto} is unreliable \cite{Kallosh:2018wme,Cicoli:2018kdo}. The 10d results of~\cite{Moritz:2017xto} were supposed to be supported by the 4d calculations, but the results of the 4d calculations were quite the opposite.

First of all, as shown in~\cite{Kallosh:2018wme}, the nilpotency condition is not satisfied in the 4d model for the parameters considered in~\cite{Moritz:2017xto}, so the 4d model of~\cite{Moritz:2017xto} is not internally consistent. An attempt to resolve this problem has been made in \cite{Moritz:2018ani}, but the new model proposed in~\cite{Moritz:2018ani} still suffers from similar problems for some values of its parameters. A set of fully consistent models of that type is constructed in~\cite{Kallosh:2018psh}. All presently available consistent 4d generalizations of the KKLT scenario, in the domain of their validity, invariably confirm the existence of dS vacua in the KKLT scenario \cite{Kallosh:2018wme,Kallosh:2018psh}.



Importantly, the authors of~\cite{Moritz:2017xto} admitted that their criticism would not apply to the version of the KKLT model \rf{adssup}, \rf{w0stab}  with a strongly stabilized dS vacuum~\cite{Kallosh:2004yh}. Because of the strong moduli stabilization, this model, and the similar models \rf{eq:K} and \rf{STU2}~\cite{Kallosh:2014oja} discussed in the previous section, are most suitable for cosmological applications.


\section{Towards full-fledged string theory solutions describing dS}\label{fullfledged}

All the models presented in \cite{Obied:2018sgi}  based on earlier constructions in Type II string theory are known as `full-fledged string theory solutions'.  They have also been more recently analyzed in \cite{Roupec:2018fsp}. These models describe classical Calabi-Yau compactifications of Type II string theory with fluxes, D-branes and O-planes, or a more general class of manifolds  with an $SU(3)$ structure. They are based on $d=10$ supergravity  with NS-NS and R-R fluxes, with D-branes and  orientifolds, and have to satisfy the tadpole and flux quantization conditions. The system is viewed without $\alpha^{\prime}$ and $g_s$ string theory corrections, which requires for consistency a large volume of compactification for the supergravity approximation to be valid, and a small string coupling. 

The meaning of these conditions in string theory is explained in detail in the review paper~\cite{Blumenhagen:2006ci} written in 2006. A specific role of tadpole conditions in the Type IIA theory was clarified in \cite{DeWolfe:2005uu}. For Type IIB on $SU(2)$-structure orientifolds,  the dictionary from string theory ingredients to \K\, potential and superpotential in standard  $\cN=1$ supergravity in $d=4$ is given in~\cite{Caviezel:2009tu}.
The full-fledged string theory solution for Calabi-Yau compactification provides a dictionary between effective low-energy $\cN=1$ supergravity with some set of chiral multiplets and the information about the fluxes and branes and orientifolds, which correspond to a specific choice of the string theory model. 

These constructions allow only $V=e^{K}\big(|D_iW|^{2}-3|W|^{2}\big)$ as consistent supergravity potentials, where only standard chiral multiplets are included.
Several years ago, this setting was full-fledged since this case covered the most general $d=4$ supergravity. However, dS supergravity~\cite{Bergshoeff:2015tra,Hasegawa:2015bza,Kallosh:2015tea,Kallosh:2015sea,Schillo:2015ssx,Kallosh:2016ndd,Ferrara:2016een,DallAgata:2016syy,Freedman:2017obq}, which we discussed in Sections~\ref{KKLT} and~\ref{nogo}, and will discuss in more detail in Appendix \ref{dSsupergravity}, has a different  potential, $V =e^{K}\big(F^2+ |D_iW|^{2}-3|W|^{2}\big)$, when an $\overline {D3}$-brane or a nilpotent multiplet is present in the theory and there is a positive term $e^{K}F^2$ absent in `full-fledged string theory solutions'. The string theory realization of the nilpotent goldstino has been proposed in~\cite{Kallosh:2015nia}. One would expect that 
the new `full-fledged string theory solutions'  will  take into account these recent developments.  

A significant progress in this direction is reported in \cite{Kallosh:2018nrk}, which goes beyond the standard uplifting by an $\overline {D3}$-brane. It is found in~\cite{Kallosh:2018nrk} that, in general, when one adds $\overline{Dp}$-branes to local sources in $d=10$, one finds $d=4$ supergravity with a nonlinear realization of supersymmetry, with chiral matter multiplets interacting with a nilpotent multiplet.  The new uplifting contribution to the supergravity potential due to  $\overline{Dp}$-branes is universal, for any  $\overline{Dp}$-branes for which the supersymmetric cycles of dimension $p-3$ are available. 

As a result, the landscape of opportunities for dS vacua  has increased dramatically. It is necessary to study these new models to find `full-fledged string theory solutions' for stable dS vacua. As of now, we have already found in \cite{Kallosh:2018nrk} dS vacua in string inspired supergravity models, which for the last decade suffered from the so-called ``obstinate tachyon'' problem. In the new context, the tachyon disappears, and a metastable dS vacuum emerges \cite{Kallosh:2018nrk}.

\section{The swampland }\label{SW}

An attempt to propose an alternative to the string theory landscape was recently made by Ooguri, Vafa et al. in  \cite{Obied:2018sgi}. The authors have suggested two new conjectures:

\begin{enumerate}

\item The first one is a no-dS conjecture, stating that a consistent theory of quantum gravity based on string theory cannot describe stable or metastable dS spaces. This conjecture has been based on various arguments and no-go theorems discussed in \cite{Brennan:2017rbf,Danielsson:2018ztv,Obied:2018sgi}. We gave a critical discussion of these arguments in the previous section, and will return to it in Appendix~\ref{dSsupergravity}.   

\item A stronger version of this conjecture is that the scalar field potential for all consistent theories should satisfy the constraint
 \be
{| \nabla_\phi V|\over V}  \geq c\, , \qquad c\sim 1 \ .
\label{swamp}\ee 
\end{enumerate}
Even though these two conjectures are related, they are partially independent. In particular, the first no-dS conjecture does not require $c \sim 1$. We analyze both of these conjectures in the present paper, as well as the proposal made in~\cite{Obied:2018sgi} for replacing the cosmological constant by string theory quintessence.

The authors of \cite{Obied:2018sgi} have been very careful in expressing their own opinion on these conjectures. For example, in the beginning of \href{https://www.youtube.com/watch?v=fU8sJRCRz24&t=1904s} {his talk at Strings 2018}, Vafa repeated, three times, that this was just a speculation, but argued that it would be interesting to entertain it nevertheless, having in mind its possible cosmological implications.  

The motivation for the conjecture \rf{swamp} has been explained as follows: 
If we assume that dS states are impossible in string theory, what could we offer as an alternative explanation for the present stage of cosmic acceleration? An often discussed possibility is that dark energy is represented by the potential of a quintessence field. Its present value should be $V\sim \mathcal{O}(10^{{-120}})$, which represents an enormous fine-tuning. This is one of the problems addressed in the context of the string theory landscape. In the theory of quintessence, this problem remained unsolved. In fact, this theory requires double fine-tuning:  in addition to the fine-tuning  $V\sim \mathcal{O}(10^{{-120}})$, one should also have $| \nabla_\phi V| \lesssim V \sim \mathcal{O}(10^{{-120}})$.

One could hope that it would be possible to reduce this double fine-tuning to the single fine-tuning $V\sim \mathcal{O}(10^{{-120}})$ by making a conjecture that it is required to have $| \nabla_\phi V|   \geq c V$ with  $c\sim 1$. But this conjecture does not help to explain why $V\sim \mathcal{O}(10^{{-120}})$,  and it does not remove the second fine-tuning $| \nabla_\phi V| \lesssim V \sim \mathcal{O}(10^{{-120}})$. Indeed, the swampland conjecture  $| \nabla_\phi V| \geq c V$ allows  {\it all} values of $|\nabla_\phi V|$ greater than $\mathcal{O}(10^{{-120}})$, which is the {\it opposite} of the quintessence requirement $| \nabla_\phi V| \lesssim V$. 
Therefore, it seems that the main goal of proposing~\rf{swamp} has been to provide some hypothesis formalizing the no-dS conjecture. From this perspective, the condition $c\sim 1$ is not required, even though it is satisfied in many string theory models discussed in~\cite{Obied:2018sgi}, which is the main reason why those models are ruled out by observations, as we show in this paper.

We explain in Sections \ref{observations}, \ref{models} and \ref{problems} why it is very difficult to overcome this problem, and point out some other problems that may plague these models. Importantly, our conclusions do not rely on  the  conjecture  \rf{swamp} with $c \sim 1$. Our results follow directly from the comparison of the predictions of the models derived from string theory, presented in~\cite{Obied:2018sgi}, with cosmological observations.

The conjecture~\rf{swamp} has been applied in~\cite{Obied:2018sgi} only to the fields  describing quintessence. One could extend it to include the Standard Model~\cite{Denef:2018etk}, inflation, etc., but such generalizations would disfavor this conjecture even more strongly. For example, the expression for the tensor to scalar ratio~$r = 8(V_{,\phi}/V)^{2}$, which is satisfied in the vast majority of inflationary models, in combination with the latest observational data \cite{Akrami:2018odb} implies that during inflation one has $| \nabla_\phi V |/V < 0.09$.  An analysis of related issues in~\cite{Agrawal:2018own,Kinney:2018nny}  gives similar constraints on $c$. The constraint $| \nabla_\phi V |/V < 0.09$ strongly disfavors the original conjecture \rf{swamp} with $c \gtrsim 1$, if applied to inflation.

However, as we have already mentioned, if the main motivation for the conjecture~\rf{swamp} has been to give a formal representation for the no-dS conjecture and possibly reduce the degree of fine-tuning in the quintessence theory, then there is no obvious reason to require $c \gtrsim 1$. 

Moreover, there is no reason to apply this conjecture  to inflationary models. Indeed, unlike the old inflationary scenario \cite{Guth:1980zm}, which assumed that inflation occurs in a metastable dS space, all realistic inflationary models are based on the slow-roll mechanism \cite{Linde:1981mu,Linde:1983gd}. The amplitude of inflationary perturbations in these models is inversely proportional to $| \nabla_\phi  V|$, so their predictions are well defined only sufficiently far away from the dS regime. Inflationary perturbations are small as long as  $| \nabla_\phi V| \gtrsim V^{3/2}$  \cite{Mukhanov:1985rz,Linde:2005ht}, and they are small enough to match the observational data if $|\nabla_\phi V| \gtrsim 10^{5 }\, V^{3/2}$ \cite{Akrami:2018odb}. This  ensures that the no-dS requirement is satisfied {\it automatically} in all slow-roll inflationary models matching the observational data  \cite{Akrami:2018odb}. An additional unmotivated constraint  on inflation of the type~\rf{swamp} would not serve any obvious purpose. Therefore, in this paper we disregard any potential implications of the conjecture~\rf{swamp} for inflation, or any arguments against the swampland conjecture \rf{swamp} based on inflation,  and, following \cite{Obied:2018sgi}, we concentrate on the theory of dark energy/quintessence.

\section{Dark energy and the cosmological data}\label{observations}

Before we continue with the implications of the swampland conjecture \rf{swamp} in the context of dark energy, let us investigate the current observational constraints on dark energy models relevant for our discussions. Particularly, in this section we focus on the `vanilla' exponential quintessence model with a potential of the form
\begin{equation}
V(\phi) = V_0\, e^{\lambda \phi},\label{eq:singexppot}
\end{equation}
where $\lambda > 0$ is a dimensionless constant. By changing the sign of $\phi$ (i.e. $\phi\to -\phi$), one can equivalently represent this potential as $V_0\, e^{-\lambda \phi}$.

This potential is interesting for two reasons. Firstly, as we see in the next sections, all the string theory based models that we consider in this paper predict a simple exponential potential or a combination of two exponentials. Additionally, as discussed in~\cite{Agrawal:2018own}, this exponential potential with a constant $\lambda$ is the \textit{least constrained} form of a quintessence potential, and by constraining it we automatically constrain more sophisticated potentials with $\phi$-dependent $\lambda$. It is also interesting to note that a constant $\lambda$ is the solution to $V_{,\phi}/V=c$ (with $c=\lambda$); cf. the swampland conjecture~\rf{swamp} for a single field $\phi$. Even though in string theory constructions this $\lambda$ is derived from the first principles, in this section we treat it as a free parameter and study the late-time observational constraints imposed on it. The exponential potential (\ref{eq:singexppot}) is a classic example of the quintessence scenario, and has been motivated and studied from the points of view of both string theory/particle physics and phenomenological approaches to dark energy; see, e.g.,~\cite{PhysRevD.32.1316,HALLIWELL1987341,Burd:1988ss,WETTERICH1988668,Wands:1993zm,Copeland:1997et,PhysRevLett.79.4740,PhysRevD.58.023503,Liddle:1998xm,Barreiro:1999zs,Heard:2002dr,Copeland:2006wr,Gupta:2011ip,Chiba:2012cb,Chang:2016aex,Hara:2017ekj,Scherrer:2018cdx}.

Our discussion in this section is restricted to the study of the background cosmological evolution, and we make use only of purely geometrical tests of the background expansion. In general, in every beyond-$\Lambda$CDM scenario, one expects interesting observable effects not only at the background level, but also at the level of the cosmological perturbations. As an example, in the presence of a nonminimal coupling of the scalar field to the matter sector one expects an enhancement of the gravitational attraction, hence a more intensive structure formation in the universe. Moreover, in many such scenarios the gravitational attraction even becomes a function of the spatial and/or temporal scales. Finally, an interesting feature of many such scenarios is that the gravitational lensing is modified, and the weak lensing measurements of galaxies can strongly constrain the models. However, in more conventional scenarios where gravity is standard and the scalar field is minimally coupled to gravity and matter, including the models we study in this paper, such modifications do not occur, hence the galaxy clustering and weak lensing measurements are not expected to introduce additional strong constraints. An important observation to make later, however, is that the constraints purely on the background dynamics of our models are so strong that they rule out all the models of interest studied in the next section, with more than $3\sigma$ confidence, even without adding extra constraints. This means that even if the additional observational data sets would introduce relevant constraints, they would not change our general conclusions here; in contrary, our conclusions would only be strengthened. 

The cosmological data sets used in our analysis consist of the Pantheon compilation of $\sim 1050$ Type Ia supernovae (SNe Ia)~\cite{Scolnic:2017caz}, the latest geometrical constraints imposed by the cosmic microwave background (CMB)~\cite{Aghanim:2018eyx} and the baryon acoustic oscillations (BAO)~\cite{Alam:2016hwk}, and the local measurements of $H_0$, the present value of the Hubble function~\cite{Riess:2016jrr}. For the SNe Ia data, we make use of its equivalent, compressed form provided in~\cite{Riess:2017lxs}, where the information of all the Pantheon supernovae is encoded into constraints on the function $E(z) \equiv H(z)/H(z = 0)$ at 6 different redshifts; this data set contains information from 15 additional SNe Ia at redshift $z > 1$. Throughout our analysis we assume a flat universe, which is also the assumption made in the analysis of~\cite{Riess:2017lxs}. We should point out that although our data sets are the latest available ones, we do not include, for example, the full CMB information provided by the Planck temperature and polarization power spectra as has been used, for instance, in~\cite{Scolnic:2017caz} in order to obtain the tightest current constraints on various parametrizations of dark energy when all available cosmological data sets are combined. For that reason, the constraints we find in this work are somewhat conservative, and our bounds on, for example, the $\lambda$ parameter would be even tighter if the full CMB data were used. As we see, however, the bounds we find are already quite tight, and sufficient for excluding all the string theory based models studied in the next section.

We perform a Markov Chain Monte Carlo (MCMC) analysis of the parameter space of our exponential model~(\ref{eq:singexppot}), and derive the Bayesian constraints on the model parameters. For every point in the Markov chain we exactly solve the scalar field equation of motion together with the Friedmann equation, given, respectively, by
\begin{eqnarray}
\phi^{\prime\prime}&+&(3-\epsilon)\phi^{\prime}+\frac{1}{H^2}\frac{\dd V(\phi)}{\dd \phi}= 0 \,,\label{eq:KG}\\
H^2 &=& \frac{V(\phi)+3H_0^2 \Omega_\text{M} e^{-3N}+3H_0^2  \Omega_\text{R} e^{-4N}}{3-\frac{1}{2}{\phi^\prime}^2} \,,\label{eq:Friedmann}
\end{eqnarray}
where a prime denotes a derivative with respect to the number of $e$-foldings $N\equiv\ln{a}$ (with $a$ being the scale factor), and the slow-roll parameter $\epsilon$ is given by
\begin{equation}
\epsilon\equiv -\frac{H^\prime}{H}=\frac{1}{2}{\phi^\prime}^2+\frac{1}{2}\frac{H_0^2}{H^2}(3\Omega_\text{M} e^{-3N}+4\Omega_\text{R} e^{-4N}) \,.\label{eq:epsilon}
\end{equation}
As usual, $\Omega_\text{M}$ and $\Omega_\text{R}$ are the present-day fractional energy densities of matter and radiation, respectively, and $H\equiv \dot a /a$ is the Hubble expansion rate with the value of $H_0$ today.

The scalar field (dark energy) equation of state is given by
\begin{equation}\label{eq:wDEdef}
w_\text{DE}=\frac{\frac{1}{2}{\phi^\prime}^2H^2-V(\phi)}{\frac{1}{2}{\phi^\prime}^2H^2+V(\phi)} \,,
\end{equation}
while the effective (or total) equation of state $w_\text{eff}$ is given by
\begin{equation}\label{eq:weff}
w_\text{eff}=-1+\frac{2}{3}\epsilon \,.
\end{equation}
During the radiation and matter domination epochs, $w_\text{eff}$ is $1/3$ and $0$, respectively, corresponding to $\epsilon=2$ and $\epsilon=3/2$.

Our cosmological model contains five free parameters, $V_0$, $\lambda$, $\phi_0$, $\Omega_\text{M}$ and $\Omega_\text{R}$ (as far as the cosmological background dynamics are concerned). Here $\phi_0$ is the initial value of the scalar field, which we set at a time well inside the radiation domination epoch ($N\sim -15$). We also assume that the field is initially at rest, i.e. $\phi^{\prime}_0=0$. It is important to note that the structure of the model implies one of the two parameters $V_0$ and $\phi_0$ to be redundant, and by assuming a sufficiently wide scanning range for one of them, we can fix the other to any specific value.
\begin{figure}[t!]
\begin{center}
\includegraphics[scale=0.52]{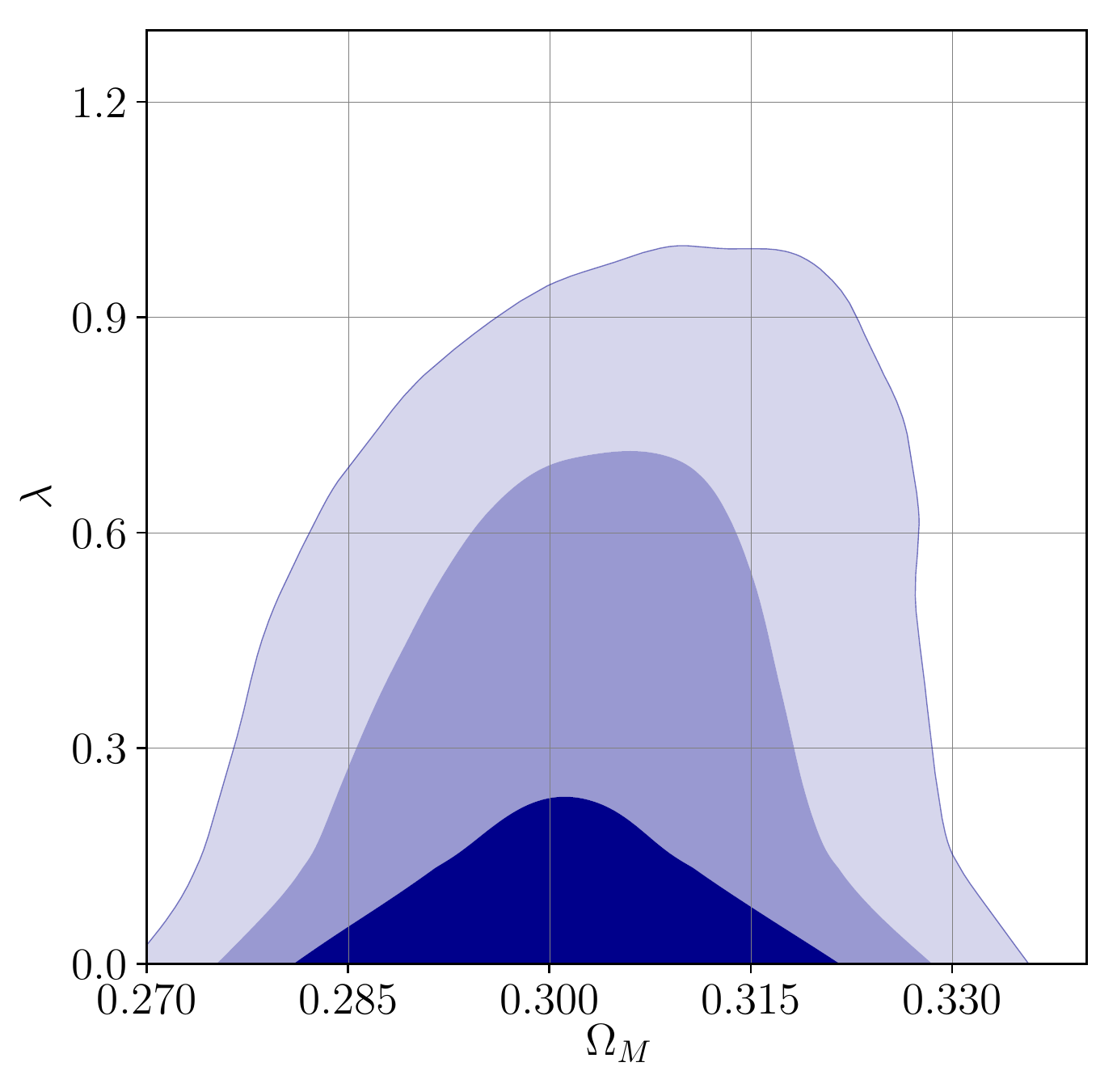}
\includegraphics[scale=0.52]{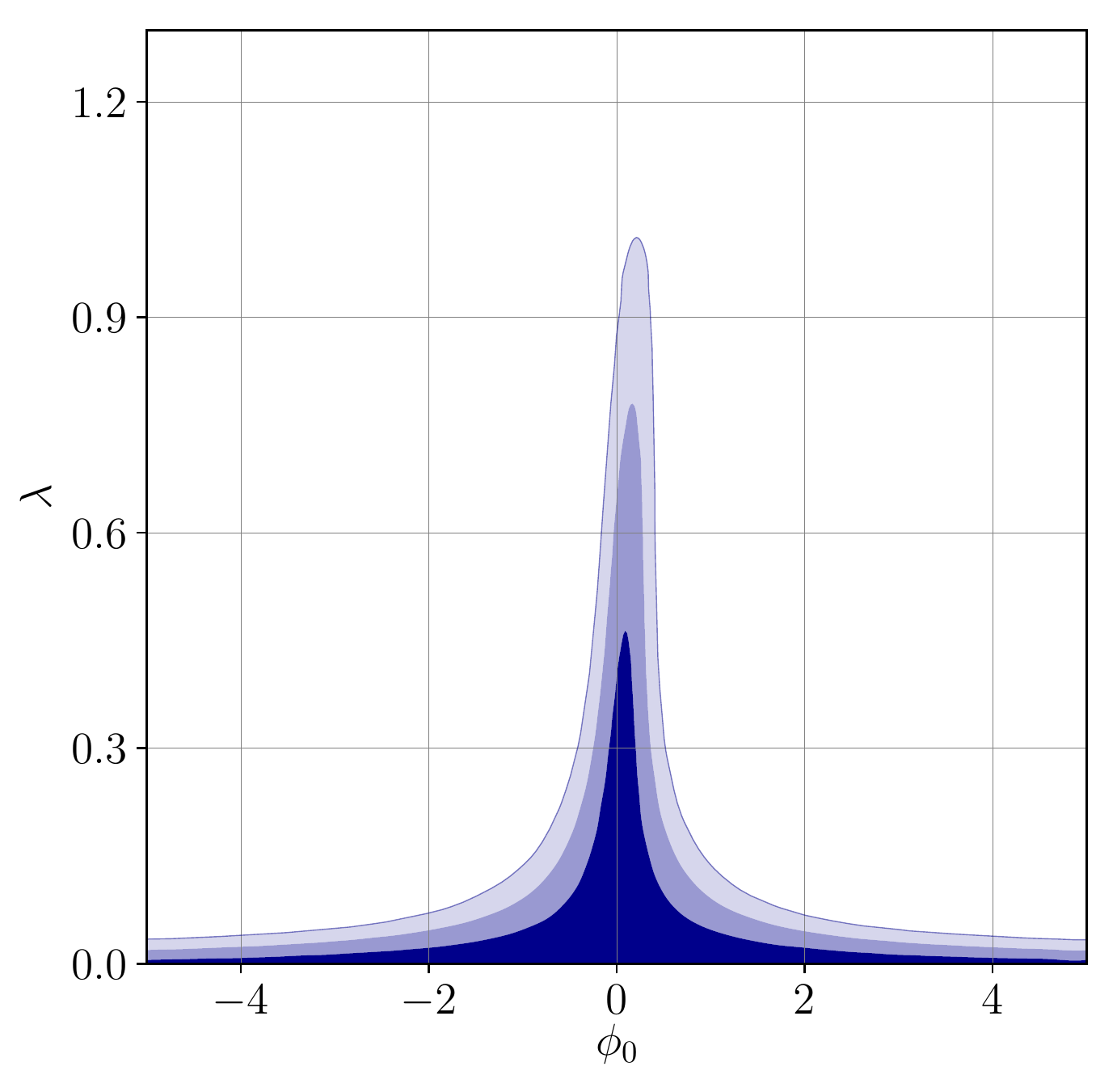}
\end{center}
\caption{\footnotesize Two-dimensional, marginalized constraints on $\lambda$ versus $\Omega_\text{M}$ (left panel) and $\lambda$ versus $\phi_0$ (right panel) for the quintessence model with the exponential potential $V(\phi) = V_0 e^{\lambda \phi}$. The contours show $68\%$, $95\%$ and $99.7\%$ confidence levels. Here, we have fixed the parameter $V_0$ to $0.7(3H_0^2)$ and varied the other parameters of the model, i.e. $\lambda$ and $\phi_0$, as well as $\Omega_\text{M}$. The one-dimensional, marginalized upper bounds on $\lambda$ are $\sim0.13$, $\sim0.54$ and $\sim0.87$, with $68\%$, $95\%$ and $99.7\%$ confidence, respectively.}
\label{fig:mainConstraints_V0p7}
\end{figure}

Fig.~\ref{fig:mainConstraints_V0p7} shows the results of our MCMC scan for the case in which $V_0$ has been fixed to $0.7(3H_0^2)$, where $3H_0^2\sim 10^{-120}$ is the critical density today, and we have scanned over the rest of the parameter space, including $\phi_0$. Here, flat priors have been imposed on the free parameters. The plots show the two-dimensional $68\%$, $95\%$ and $99.7\%$ confidence regions in the $\lambda-\Omega_\text{M}$ and $\lambda-\phi_0$ planes. Marginalizing the full posterior probability density function over all the parameters except $\lambda$, we obtain the upper bounds of $\sim0.13$, $\sim0.54$ and $\sim0.87$ on $\lambda$ with $68\%$, $95\%$ and $99.7\%$ confidence, respectively. An interesting observation from the right panel of Fig.~\ref{fig:mainConstraints_V0p7} is the rapid drop of $\lambda$ by increasing $|\phi_0|$. The contours peak at $\phi_0\sim 0.4$ and then quickly decrease when $\phi_0$ deviates from the peak value.

Even though, as we mentioned above, one of the two parameters $V_0$ and $\phi_0$ is redundant, and we have therefore fixed $V_0$ and varied $\phi_0$, this redundancy holds only when $\lambda$ is strictly nonzero. This exceptional case corresponds to a constant dark energy, i.e. a cosmological constant with $\Lambda=V_0$, independently of the value of $\phi_0$. Since we have chosen $V_0$ to be $0.7(3H_0^2)$, the only $\Lambda$CDM case that we have in our parameter space is with $\Omega_\Lambda=0.7$. Even though this value is consistent with the measured value of $\Omega_\Lambda$ for $\Lambda$CDM, the observational uncertainties have not been taken into account, and our results are, statistically speaking, not complete for the $\Lambda$CDM corner of the parameter space. This may slightly bias the constraints on $\lambda$.

\begin{figure}[t!]
\begin{center}
\includegraphics[scale=0.52]{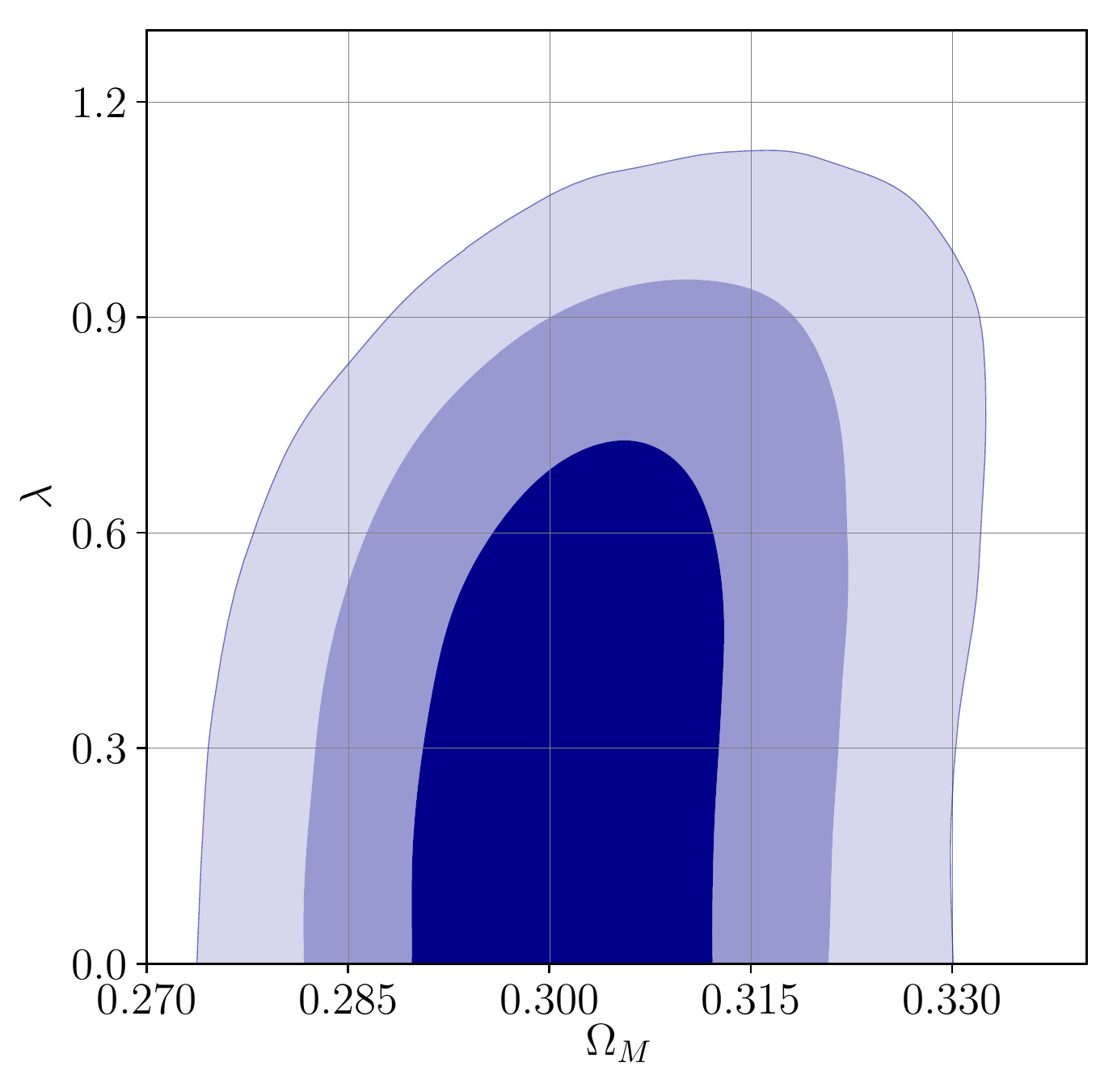}
\end{center}
\caption{\footnotesize The same as in Fig.~\ref{fig:mainConstraints_V0p7}, but when $\phi_0$ is fixed to $0$ and $V_0$ is varied. The marginalized, one-dimensional $68\%$, $95\%$ and $99.7\%$ upper bounds on $\lambda$ in this case are $\sim0.49$, $\sim0.80$ and $\sim1.02$, respectively. These results are in excellent agreement with our findings based on a frequentist, profile likelihood analysis, demonstrating that they are the least prior-dependent results we can obtain from an MCMC-based, Bayesian analysis, and provide the weakest possible bounds on $\lambda$.}
\label{fig:mainConstraints_phi00}
\end{figure}

For this reason, we have also performed a statistical analysis when $V_0$ has been allowed to vary as well. Our results show that the bounds on $\lambda$ do not change significantly, as long as the allowed range of $\phi_0$ is not too large. Enlarging the range of $\phi_0$ increases the volume of the parameter space (mostly) around $\lambda=0$, and therefore increases the probability of the model to give a $\Lambda$CDM-like cosmology (which provides a good fit to the data). This in turn biases our results towards $\Lambda$CDM (i.e. small $\lambda$) and affects the marginalized upper bound on $\lambda$ by lowering it to smaller and smaller values, as confirmed by our statistical results; this is a consequence of our Bayesian framework, where priors may play an important role in situations like ours here. The weakest bound on $\lambda$ is therefore expected when $\phi_0$ is fixed to a specific value and $V_0$ is varied.\footnote{Note that we can still explore the entire parameter space with this parametrization, while the effects of priors related to the range of $\phi_0$ are minimized.} We show in Fig.~\ref{fig:mainConstraints_phi00} the two-dimensional contour plot in the $\lambda-\Omega_\text{M}$ plane for a scan with $\phi_0$ having been fixed to $0$.\footnote{It is important to note that there is additionally some small dependence of the $\lambda$ bounds on the \textit{effective} priors imposed upon the parameters in the MCMC process. Since $\lambda$ and $\phi_0$ sit in the exponent of the potential, flat priors on these parameters impose an effective non-flat prior on the combination $V_0 e^{\lambda \phi_0}$, which then translates into an effective non-flat weighting of the $\lambda$ parameter itself. Our tests show that this prior effect is larger when $\phi_0$ is fixed to a nonzero value, as expected. We therefore fix $\phi_0$ to zero in order to minimize this additional prior effect on $\lambda$ as well.} The figure shows a slight increase in the bounds on $\lambda$. The marginalized, one-dimensional $68\%$, $95\%$ and $99.7\%$ upper bounds on $\lambda$ in this case are $\sim0.49$, $\sim0.80$ and $\sim1.02$, respectively. In spite of these small changes of the bounds depending on which exact priors and ranges one imposes on the parameters, our results show that one never obtains a $3\sigma$ bound on $\lambda$ larger than $\sim 1$.\footnote{In order to directly see that this $\lambda \lesssim 1$ is the least tight $3\sigma$ constraint on $\lambda$, we additionally performed a {\it profile likelihood} analysis of the parameter space; see, e.g.,~\cite{Rolke:2004mj,Akrami:2009hp,Akrami:2010cz} and references therein. This is a \textit{frequentist} approach, where the statistical results are independent of priors and ranges. The contours and the upper bounds on $\lambda$ that we obtained through the profile likelihood analysis were almost identical to what we have found in our Bayesian analysis of Fig.~\ref{fig:mainConstraints_phi00}, demonstrating that they are the least prior-dependent results we can get from an MCMC-based Bayesian analysis. It also confirms that $\lambda$ cannot be larger than $\sim 1$ under any circumstances, with more than $3\sigma$ confidence.} By trying to be as ignorant and unprejudiced as possible about the values of the parameters before comparing the model to the data through enlarging the ranges of the parameters in our MCMC scans, especially for the initial value $\phi_0$, this bound on $\lambda$ does become even tighter. We see in the next section that this $3\sigma$ upper bound of $\sim1$ rules out all the models considered in~\cite{Obied:2018sgi}.

It is important to note that the statistical constraints on $\lambda$ obtained above ignore the issue of the probability to begin the last stage of the cosmological evolution in an immediate vicinity of the point $\phi_{0} $ very close to the very narrow peak at $\phi_{0} \sim 0.4$ shown in the right panel of Fig. \ref{fig:mainConstraints_V0p7}. For any other initial conditions, the probability to describe the present state of the universe in models with $\lambda \sim 1$ is vanishingly small.

Bounds on $\lambda$ for the same single-exponential potential~(\ref{eq:singexppot}) have been provided also in the two recent papers~\cite{Agrawal:2018own,Heisenberg:2018yae} on the swampland conjectures. The results of these papers are not based on a rigorous statistical analysis of the model, and our findings are  in strong disagreement with the work of Heisenberg et al.~\cite{Heisenberg:2018yae}. For that reason, we dedicate Appendix~\ref{sec:priorbounds} to a detailed comparison of our results and methodology with those of~\cite{Agrawal:2018own,Heisenberg:2018yae}.  

\

\

{\bf{The case of double-exponential potentials:}} As we see in the next section, there are string theory based models of quintessence with potentials that are of the form
\begin{equation}
V(\phi) = V_1 e^{\lambda_1 \phi} + V_2 e^{\lambda_2 \phi} \,.\label{eq:2exppotA}
\end{equation}
We study these double-exponential models and compare them with the cosmological data in Appendix~\ref{sec:doubleexp}.
One important result of this investigation is that the constraint on the smallest of the two exponent coefficients $\lambda_1$ and $\lambda_2$ is nearly identical to the constraint on the single exponent coefficient $\lambda$ studied above. A similar conclusion is valid for the models $V(\phi) = V_1 e^{-\lambda_1 \phi} + V_2 e^{-\lambda_2 \phi}$, since they are equivalent to the models~\rf{eq:2exppotA} when $\phi \to -\phi$.

\section{Accelerating universe according to the swampland conjectures }\label{models}

In this section, we discuss the string theory models of accelerating universe described in~\cite{Obied:2018sgi,Agrawal:2018own}. Our investigation of these models does not require any use of the no-dS conjecture and the constraint~\rf{swamp}. This discussion can be applied to both dark energy/quintessence and inflation described by such models, with some caveats. 

Inflation is a stage of quasi-exponential expansion in the early universe, with a Hubble rate $H$ which can be of the order of $10^{-5} $ in Planck units, whereas at the present stage of the acceleration of the universe one has $H \sim 10^{{-60}}$. The difference in scales is colossal, but many of our conclusions depend only on the scale-independent  ratio ${| \nabla_\phi V|\over V}$. 
On the other hand, in the discussion of dark energy the main emphasis is on whether the acceleration may occur now, rather than how it may end. Meanwhile in the discussion of inflation, we must also study how exactly it ends, how the universe reheats after that, etc. Observational constraints on inflation are much more stringent than those on dark energy. Therefore, the general expectation is that if the models we discuss here cannot describe dark energy, they cannot describe inflation either. We will return to this comment later on. 

There are models in~\cite{Obied:2018sgi} where the value of the constant $c$ in eq.~\rf{swamp} is given for spaces with dimensions (after compactification) different from $d=4$, in particular for $d=10$ and $d= 5$. However, comparing theoretical models with observations makes sense only for $d=4$. Therefore, here we consider only the cases where the value of $c$ has been given for $d=4$.

\subsection{M-theory compactifications} \label{mm}

The first example in~\cite{Obied:2018sgi}, based on the hyperbolic compactification of 11-dimensional supergravity/M-theory with fluxes, has a potential depending on two exponential functions of the canonical scalar field $\phi$,\footnote{Note again that by taking $\phi\to -\phi$ one can equivalently work with a term of the form $V_0\, e^{-\lambda \phi}$ in the potential instead of $V_0\, e^{\lambda \phi}$, with the same $\lambda>0$. Even though we chose the convention of writing the potentials as $V_0\, e^{\lambda \phi}$ in Section~\ref{observations}, Appendix~\ref{sec:doubleexp} and many other places in the present paper, we use the opposite convention of $V_0\, e^{-\lambda \phi}$ in this section in order to be consistent with the notations adopted in~\cite{Obied:2018sgi}.}
\be\label{ob}
V= V_{\mathcal{R}} e^{- \sqrt{18\over 7} \phi} +  V_{G} e^{- { \sqrt {50\over 7}}\phi} \ .
\ee
The first term is due to a negative curvature of the compactified space, and the origin of the coefficient $\sqrt{18\over 7}$ in the first exponent is\footnote{Here, we denote the larger coefficient by $\lambda_1$ and the smaller one by $\lambda_2$, in order to be consistent with the notations of Appendix~\ref{sec:doubleexp}.}
\be
\lambda_2={6\over \sqrt{(d-2) (d_\text{cr}-d)}}\Big |_{d=4, d_\text{cr}=11}= \sqrt{18\over 7} \approx 1.6 \ .
\label{Mcase}\ee
The second term is due to fluxes.
In fact, an M-theory model of accelerating universe with a very similar potential,
\be\label{gutp}
V= V_{\mathcal{R}} e^{- \sqrt{18\over 7} \phi} +  \tilde V_{G} e^{- { \sqrt {14}}\phi} \,,
\ee
was already studied 15 years ago  in \cite{Gutperle:2003kc,Emparan:2003gg}. In both models, the flux-type exponential with $\lambda_1=\sqrt{50\over 7}$ or $\lambda_1=\sqrt{14}$ is too steep for describing dark energy at late times. Therefore,
the only possibility to have a reasonable late-time cosmology is in a regime with large $\phi$, where the second exponential is small.
In this regime, which is equivalent to a negligible 4-form field contribution, the model~\rf{ob} coincides with the model~\rf{gutp} studied in~\cite{Gutperle:2003kc,Emparan:2003gg}.  
The exponential potential of dark energy today, neglecting fluxes, is therefore of the form
\be
V_\text{DE}\approx V_{\mathcal{R}} e^{- \sqrt{18\over 7}\phi} \sim 10^{-120}. 
\ee
This model was marginally consistent with the dark energy data in 2003, when $\la_{2}\sim 1.7$ was still in agreement with the data, but required some fine-tuned initial conditions~\cite{Gutperle:2003kc,Emparan:2003gg}. 

However, taking into account the bounds on $\lambda$ for single-exponential potentials obtained in the previous section based on the current constraints on dark energy, we can conclude that this model with $\lambda_{2} \approx 1.6$ is inconsistent with the current observations with more than $99.7\%$ confidence. In Appendix~\ref{sec:doubleexp}, we have provided for interested readers a more general approach to double-exponential potentials, including a detailed statistical analysis of their parameter space. The results of Appendix~\ref{sec:doubleexp} show explicitly that the models~\rf{ob} and~\rf{gutp} are both ruled out; see Fig.~\ref{fig:double_exp_lambda1-2}

Before we look into the other models proposed in~\cite{Obied:2018sgi}, it is instructive to discuss some issues with models of accelerating universe, which are present here, independently of the statistical disagreement with the data. This will strengthen our general conclusion that the models of accelerating cosmology in  \cite{Obied:2018sgi}
tend to be in conflict  with $d=4$ general relativity.

In~\cite{Gutperle:2003kc}, the value of $V_{\mathcal{R}}$ was computed via the curvature of an internal compact space, $R_{ab} = - 6 g_{ab} {1\over r_c^2}$, and found to be
$
V_{\mathcal{R}} = - 2 R = {21\over r_c^2}
$,
so that
\be
V_\text{DE}\approx {21\over r_c^2} e^{-c\phi} \sim 10^{-120}. 
\ee 
It was found there, see also  \cite{Kaloper:2000jb,Emparan:2003gg},  that the model had extra light Kaluza-Klein (KK) modes with the Compton wavelength of the same order as the size of the observable part of the universe,
 \be
 m_\text{KK} = \mathcal{O}(e^{-c\phi/2}/r_c) \sim \sqrt{V_\text{DE}}  \sim H_{\rm DE}\sim 10^{-60} \ .
\ee
In other words, in the absence of moduli stabilization, the compactified space may effectively decompactify. This is still an open problem and remains to be solved, and 
therefore, the effective M-theory dark energy models of accelerating universe have practically been abandoned; see, e.g., a discussion of this model on page 40 of the dark energy review \cite{Copeland:2006wr}. As we see in the next subsection, the second model discussed in  \cite{Obied:2018sgi} faces a similar problem. Of course, this may not be very important since both of these models are ruled out by observations anyway, as we find $c \sim 1.6$ in both cases. Nevertheless, this issue requires careful consideration as it might be systemic in models with non-stabilized extra dimensions.

\subsection{$O(16)\times O(16)$ heterotic string}\label{O16}
This is a non-supersymmetric model without tachyons in $d=10$, which was invented in 1986 \cite{AlvarezGaume:1986jb,Dixon:1986iz}.
The dark energy potential in this model has two exponential terms, which depend on the dilaton and the volume modulus, both evolving. In terms of the two canonical fields $\hat \rho$ and $\hat \tau$, the dark energy potential is
\be
V_\text{DE}= V_{\mathcal{R}} e^{- {2\over \sqrt{d_\text{cr}- 4}} \hat \rho} e^{\sqrt{2} \hat \tau} +  V_{\Lambda} e^{\sqrt{d_\text{cr} -4} \hat \rho} e^{ 2{ \sqrt {2}}\hat \tau} \ .
\label{Strcase}\ee
Using $d_\text{cr} = 10$, one finds
\be
V_\text{DE}|_{d_\text{cr}=10}= V_{\mathcal{R}} e^{- \sqrt{2\over 3} \hat \rho} e^{\sqrt{2} \hat \tau} +  V_{\Lambda} e^{\sqrt{6} \hat \rho} e^{ 2{ \sqrt {2}}\hat \tau} \ . 
\label{Strcase2}\ee
The values of $V_{\mathcal{R}}$ and $V_{\Lambda}$ are unspecified, but somehow related to the geometry of the internal manifold and  $d=10$ cosmological constant. Today, the fields have to take values such that $V_\text{DE}\sim 10^{-120}$. The volume of the compactified manifold is proportional to $e^{6 \hat \rho}$.

\begin{figure}[t!]
\begin{center}
\includegraphics[scale=0.15]{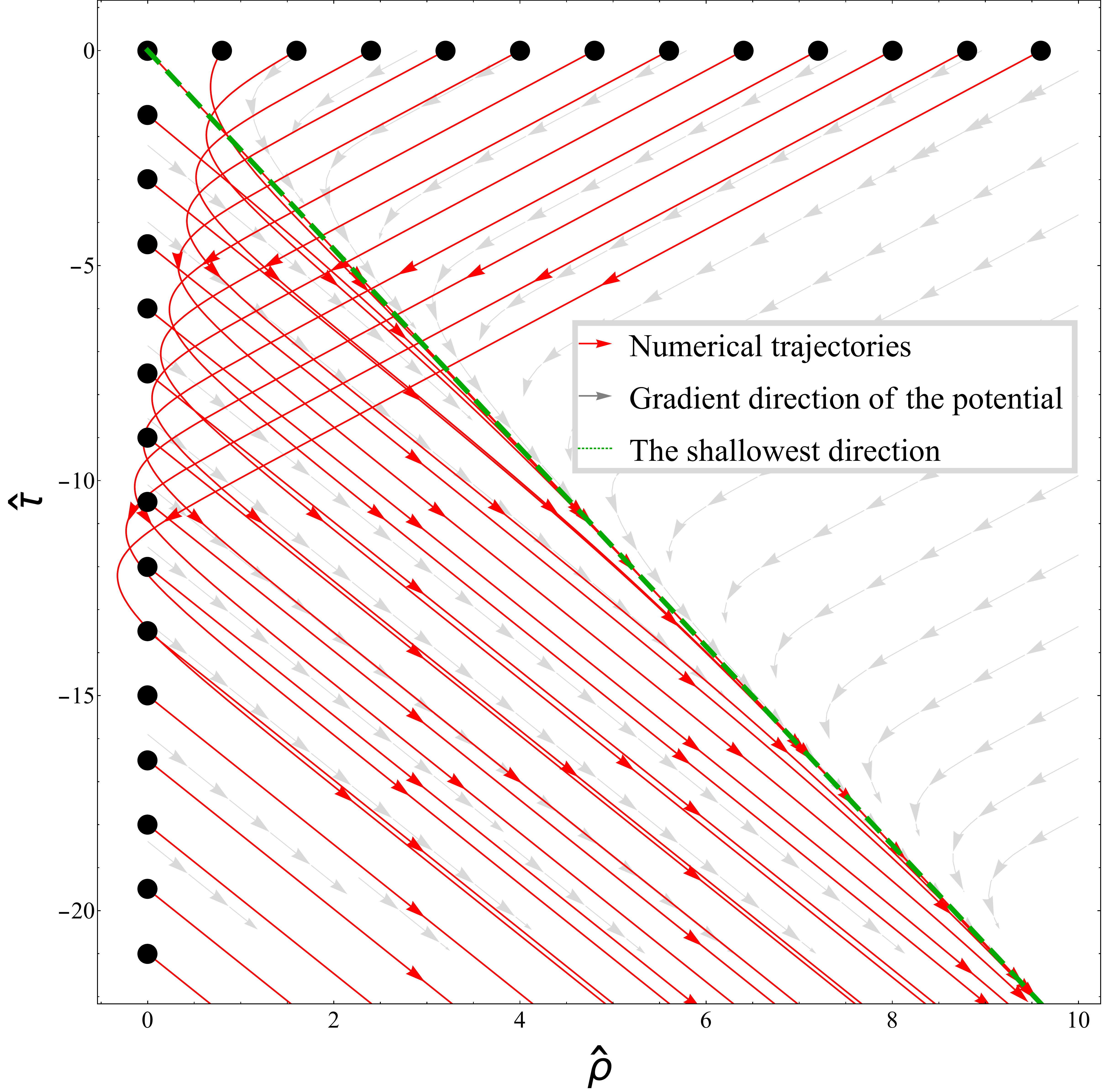}
\includegraphics[scale=0.15]{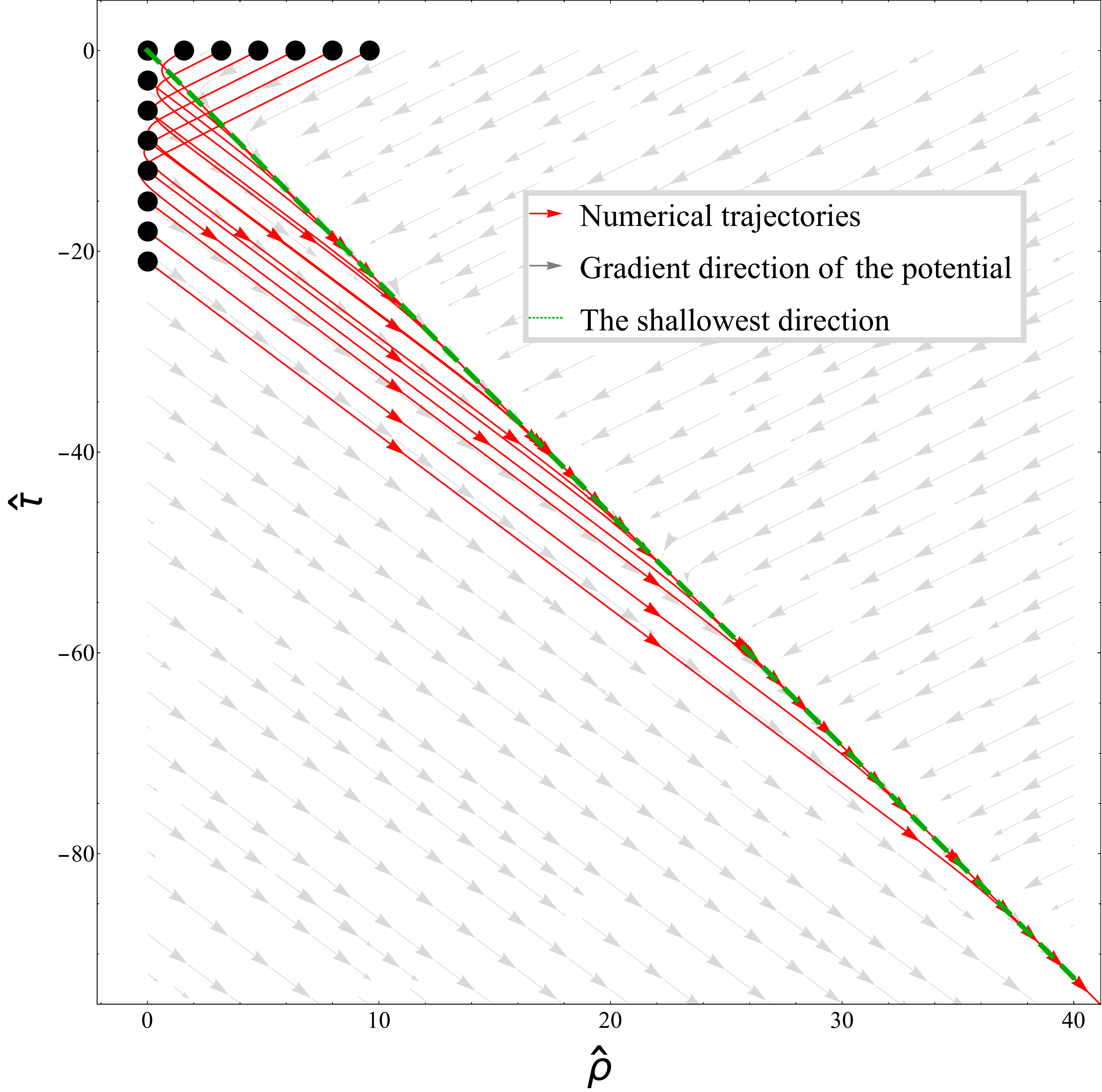}
\end{center}
\caption{\footnotesize In the left panel we show how the two-field system of Section~\ref{O16} with the scalar field potential~(\ref{Strcase2}) evolves in time, starting with different initial conditions in the $\hat \rho-\hat \tau$ plane. In the right panel we present a longer time evolution, where the system evolves in the shallowest direction (depicted by a green line). For simplicity, the evolution is shown in the absence of matter and for $V_{\mathcal{R}} = 18 V_{\Lambda}$, in which case the shallowest direction is given by $\hat \tau = -{4 \hat \rho\over \sqrt 3}$; see eq.~\rf{shall}.}
\label{fig:2fields}
\end{figure}

Since both fields are evolving, the cosmological evolution of dark energy is complicated. 
We have studied the time evolution in this model and found that independently of the initial values of the fields $\hat \rho$ and $\hat \tau$, the cosmological evolution eventually approaches the regime with the smallest value of $|\nabla_\phi V|/V$; see Fig.~\ref{fig:2fields}. In this regime 
\be\label{shall}
\hat \tau = -{4 \hat \rho\over \sqrt 3} + {1\over \sqrt 2}\log {V_{\mathcal{R}}\over 18 V_{\Lambda}} \ .
\ee
This corresponds to the smallest effective value of $c\approx  1.6$ in this model,  which is similar to the result obtained in~\cite{Obied:2018sgi}. Based on our analysis of Section~\ref{observations} we can conclude that the evolution along this shallowest direction is ruled out by the data. We study more general evolution scenarios in Appendix~\ref{sec:two-field} and conclude that this model does not exhibit cosmologically viable solutions.   

Nevertheless, it makes sense to study this model more attentively. Using the relations obtained above, one can show that if the universe evolves along the stable attractor trajectory \rf{shall} from the Planck density $V = \mathcal{O}(1)$ to the present density $\sim 10^{-120}$, the size of the compactified space during that period grows approximately $10^{29}$ times, and the volume of the compactified space increases by the factor of $10^{176}$. 
This tremendous growth of the volume of the compactified space during the cosmological evolution may strongly affect physics in the $d=4$ universe. 

This was the second model in the class of dark energy models discussed in~\cite{Obied:2018sgi}. We encountered a problem with decompactification in the first model discussed in the previous subsection~\cite{Gutperle:2003kc,Emparan:2003gg}, and now see that this second model suffers from a very similar problem. This suggests that such problems may be quite generic for  models without moduli stabilization. 

We do not explore this issue any further in this paper, simply because all the models proposed in~\cite{Obied:2018sgi} are inconsistent with observational data anyway.

\subsection{Type II string theory models}\label{t2}
The potentials discussed for the Type II string theory models of~\cite{Obied:2018sgi} depend on two moduli, dilaton and volume. They had been studied in detail in the earlier papers \cite{Hertzberg:2007wc,Caviezel:2008tf,Flauger:2008ad,Caviezel:2009tu,Wrase:2010ew}. There is a lower bound on $c$ in the IIA case of these models, which has been derived in~\cite{Hertzberg:2007wc},
  \be
  c\gtrsim 2 \ .
 \label{Shamit}  \ee
 It is, however, explained in \cite{Hertzberg:2007wc} that it is possible to evade the no-go theorem in this case, for example by considering curved compactification manifolds, instead of the flat ones, for which the bound \rf{Shamit} is valid.

Table 1 in~\cite{Obied:2018sgi} summarizes constraints on $c$ in Type IIA/B compactifications to 4 dimensions with arbitrary R-R and NS-NS flux and O$q$-planes and D$q$-branes with fixed $q$. All the cases in that table require $c\gtrsim 2$, and had been studied before in \cite{Hertzberg:2007wc,Caviezel:2008tf,Flauger:2008ad,Caviezel:2009tu,Wrase:2010ew}. In the IIB case discussed in~\cite{Caviezel:2009tu}, there is an example in a twisted tori class of models for which 
 \be
 c\gtrsim \sqrt 2 \ ,
 \ee
 and is therefore also excluded by the data.
 
There are two other cases with $c =\sqrt{2\over 3}\approx  0.8$ and $c > 1$, which belong to the ``indeterminate'' models of Table 1 in~\cite{Obied:2018sgi}. 
These models  are in the class studied in~\cite{Flauger:2008ad,Caviezel:2008tf} and the consequent papers, where it is hard to avoid the situation with $V_{,\phi}=0$. For example, in Table 1 of~\cite{Wrase:2010ew} a no-go case is presented with $\epsilon ={1\over 2} c^2 \geq  {1\over 3}$, which means $c \approx 0.8$. However, in the next columns of that paper one can see that adding $F_0$ or $F_2$ fluxes removes the no-go case and allows to find $V_{,\phi}=0$. Although it may not be a stable minimum, it  disproves the conjecture~\rf{swamp}. The second case in the group of ``indeterminate'' models is presented in Appendix B of \cite{Obied:2018sgi}, and requires $c > 1$. The authors  of \cite{Obied:2018sgi}   notice, however, that this bound is not necessarily realized since there is no string theory construction supporting such a $c$. To claim that the model is derived from string theory means that one has to present ingredients which satisfy 
consistency conditions. These include the tadpole condition, flux quantization, large volume and small string coupling requirements. No  models with $c<\sqrt 2$ satisfying all these requirements are presented   
 in  \cite{Obied:2018sgi}.

\subsection{NEC bound}

One more example in Section 2.4 of~\cite{Obied:2018sgi} is based on the null energy condition (NEC) bound, 
\be
c= \la_\text{NEC} =  \sqrt{2(d-4)\over (d-2)} \ .
\ee
For M-theory with $d=11$, we find $\la_\text{NEC} =  \sqrt{2(d-4)\over (d-2)}= \sqrt{14/9} \approx 1.25$. For superstring theory with $d=10$, we find $\la_\text{NEC} =  \sqrt{2(d-4)\over (d-2)}= \sqrt{3/2} \approx 1.22$. Since these models are not supported by any known string theory constructions, and are ruled out observationally anyway, we do not discuss them here.
\

\

\noindent In conclusion, all string theory examples which correspond to the specific string theory constructions of~\cite{Obied:2018sgi} require $c\gtrsim \sqrt 2 $, and all of them are therefore ruled out by the cosmological data, independently of the additional conceptual issues associated with many of these models.

\subsection{More examples}

In~\cite{Agrawal:2018own}, seven more references, [30-36], have been added as models which `made an attempt to embed quintessence in string theory'. We discuss those models here for completeness. Two of the references from 2001-2003 study string theory and quintessence  with examples of  exponential potentials. It is concluded that
\be
c > \sqrt 6,
\ee
which means that the models are ruled out by the data and there is no need to discuss them here.

Let us  look more carefully at the other five models, dating from 1999 to 2012, where  inflation models were converted into quintessence models. These are natural inflation models, axion monodromy  models and poly-instanton inflation models, consistent with the data on inflation. They all have exit from inflation, which means a minimum of the potential with $V_{,\phi}=0$, and therefore they do not support the conjecture \rf{swamp} in \cite{Obied:2018sgi}.

The model in \cite{Choi:1999wv} is the early string theory axion quintessence model of 1999,  where the axion is a partner of the volume modulus. The volume modulus is stabilized by some  supergravity type stabilization mechanism which was known at that time. The model can be viewed as one of the possible axion inflation models, namely,  natural inflation,  converted into quintessence.  The second model in \cite{Panda:2010uq} follows an analogous pattern. It takes the axion monodromy inflationary model with a linear potential \cite{McAllister:2008hb} and converts it into a quintessence. It is important here to note that the axion monodromy inflationary model with a linear potential \cite{McAllister:2008hb} involves KKLT or KL stabilization for consistency. Finally, the third model in \cite{Cicoli:2012tz} is also based on a particular string theory inflationary model known as `poly-instanton' inflation. This model is based on Large Volume Compactification and also gives an example of a string inflation model converted into quintessence models.
All three classes of string theory quintessence models  in \cite{Choi:1999wv}, \cite{Panda:2010uq},  \cite{Cicoli:2012tz} do not support the swampland conjecture \rf{swamp} since they are based on various constructions of moduli stabilization.

\section{Conceptual problems with string theory quintessence}\label{problems}
  \subsection{Quantum corrections}
  
In the previous sections, we studied string theory models of quintessence and compared them to the cosmological data assuming an exponential potential $V\sim e^{\la \phi}$ as a proxy for models supporting the $V_{,\phi}/V \sim c$ conjecture for $\la=c$. We compared the $V\sim e^{\la \phi}$ quintessence model to the data, and our conclusion was that models with $ c \gtrsim1.02$ were ruled out by the data at the $99.7\%$ confidence level.
 
An obvious question here is the following. Once the model $V\sim e^{\la \phi}$ with a given $\la=c$ is viewed as a legitimate string theory model, one may wonder what will happen with this model when possible quantum corrections of various kinds are taken into account. And since the scale of the potential is $10^{-120}$, one would expect that quantum corrections, for example from the Standard Model, may change the model in a way which cannot be predicted. It is stressed in~\cite{Dodelson:2013iba} that this limitation is another reason why it is not possible, based on our current knowledge, to make a robust prediction for $w$ in string theory. On the other hand, the idea of the string landscape, as depicted in Fig.~\ref{chi}, is that after taking into account quantum corrections many values of a small cosmological constant are possible. This is opposite to an attempt to protect any given model from quantum corrections of any kind, string theory or Standard Model corrections.
 
\subsection{Decompactification}
  
In the absence of moduli stabilization, one should always check whether a model really describes compactification. As we already mentioned, the first model proposed in \cite{Obied:2018sgi} practically coincides with the model studied long time ago in  \cite{Gutperle:2003kc,Emparan:2003gg}. It was found there that it did not really describe compactification. This issue is discussed in Section  \ref{mm}. Similarly, in Section \ref{O16} we found that in the second class of  models studied in  \cite{Obied:2018sgi}, the volume of the compactified manifold grows by a factor of $\sim 10^{175}$ during the cosmological evolution. Thus, this may be a systemic problem of the models without moduli stabilization. 

One may try to solve these problems,  or even use them constructively for providing an anthropic solution to the cosmological constant problem. This speculative possibility had been proposed in~\cite{Gutperle:2003kc}, but so far there has been no progress in this direction. This is not surprising though, especially having in mind that all string quintessence models studied so far are ruled out by observational data. 

\subsection{The fifth force}
  
The light quintessence scalar fields have a Compton wavelength comparable to the size of the cosmological horizon. Since they are extremely light and have a geometric origin, they may lead to a fifth force violating the equivalence principle, which has been tested with ever increasing precision. For example, the MICROSCOPE satellite mission has already confirmed the equivalence principle with an accuracy better than $10^{{-14}}$~\cite{Touboul:2017grn,Berge:2017ovy}, and the plan is to reach the level of $10^{{-15}}$. The Galileo Galilei (GG) proposal aims at increasing the precision to $10^{{-17}}$ \cite{Nobili:2018eym}.

From the point of view of those who believe in the weak gravity conjecture, this precision may not be even needed. This conjecture states that all forces must be either stronger than gravity, or vanish. The tests of the equivalence principle imply that the fifth force must be many orders of magnitude weaker than gravity. Meanwhile, the weak gravity conjecture insists that there are no interactions weaker than gravity. Therefore, the weak gravity conjecture  suggests that the fifth force due to quintessence fields acting on particles from the Standard Model must completely vanish~\cite{Brennan:2017rbf,Agrawal:2018own}. In supergravity, one can impose certain conditions on the \K\ potential and superpotential which may lead to the strong suppression or even vanishing of the fifth force \cite{Brax:2009kd,Akrami:2017cir}. Constructing realistic  models of this type in string theory without vacuum stabilization is a very challenging task. 

\subsection{Contribution of matter to the dark energy potential}

The easiest way (for us) to explain yet another issue is to remind the readers of what happens when we consider inflation in theories with the KKLT construction~\cite{Kachru:2003aw}, where the `uplift' from AdS to dS  is provided by an $\overline {D3}$-brane. In the language of a 4d effective action, this scenario can be described by the theory with the \K\ potential and superpotential \rf{KKLT2003},
\be
\begin{aligned}\label{KKLTpot}
&K= -3\log\left(T+\bar T\right)+S \bar S \, , \\
 &W= W_0 + A e^{-a T} + \mu^2 S\,,
\end{aligned}
\ee
where $T$ is the volume modulus and $S$ is a nilpotent chiral superfield. 

If we ignore the field $S$ in this model, the model would describe a theory with a potential having a minimum corresponding to a stable AdS vacuum with a negative cosmological constant. On the other hand, if we ignore the field $T$, the theory would describe a dS space with the cosmological constant $\Lambda = \mu^{4}$ provided by an $\overline {D3}$-brane.  

One could expect that when we combine these two ingredients into the KKLT model \rf{KKLTpot},  the cosmological constant  $\Lambda = \mu^{4}$ will be added to the AdS vacuum energy of the theory describing the field $T$, thus providing the required uplifting. However, the situation is more complicated: At the moment when we unify these two theories, the uplifting term $\mu^{4}$ becomes multiplied by $e^{K}$, where the total \K\ potential $K$ now includes the \K\ potential of the volume modulus $-3\log\left(T+\bar T\right)$. This produces the uplifting term proportional to ${\mu^{4}\over (T+\bar T)^{3}}$. In terms of a canonically normalized volume modulus field $\phi$, this term is not a constant, but a steep exponential potential $\sim e^{-\sqrt 6 \phi} $, rapidly falling at large values of the volume modulus. 

If this term is small, it leads, as expected, to a gradual uplifting of the AdS vacuum to a metastable  dS vacuum. But if the constant $\mu^{2}$ is too large, this new term destabilizes the volume modulus, and the field starts moving down in a steep exponential potential \cite{Kallosh:2004yh}. 
The main lesson is that if we try to add a cosmological constant in $d =10$, then in $d=4$ it acts as an exponential potential, which tries to decompactify extra dimensions.  

A similar effect may occur if one adds dark matter, or hot ultra-relativistic gas. That is why even after the KKLT potential stabilizes the volume modulus $T$, it is necessary to make sure that the contribution of other fields, including the inflationary potential \cite{Kallosh:2004yh}, radiation~\cite{Buchmuller:2004tz} and dark matter, does not destabilize it. The simplest method to do that has been proposed in~\cite{Kallosh:2004yh,BlancoPillado:2005fn,Kallosh:2007wm,Kallosh:2011qk,Dudas:2012wi}.

Similar exponential terms may appear in the non-stabilized dark energy models based on string theory~\cite{Obied:2018sgi} when one takes into account the contribution of dark matter and radiation to the volume modulus potential. This effect is well known to those who study inflation in string theory, and has been discussed in the quintessence literature as well, in the context of ``coupled quintessence'' or ``interacting dark energy''~\cite{Amendola:1999er,Barros:2018efl}. We did not include it in our analysis of the observational constraints on the exponential potential, simply because we should  first learn how to add matter to quintessence models in string theory without being in conflict with the fifth force problem discussed above. We believe that the possible contribution of dark matter and radiation to the  exponential potential of the volume modulus can only result in strengthening our constraints on the parameters of such models.

\subsection{Quintessence and the bound on field excursions}

Suppose for a moment that the quintessence potential is given by a single exponential $V\sim e^{\lambda\phi}$, and the cosmological evolution began at the Planck density with  $
V\sim e^{\lambda\phi_{0}} = \mathcal{O}(1)$. Eventually dark energy becomes small, with $V\sim e^{\lambda\phi} = 10^{{-120}} \sim  e^{-276}$. 

For definiteness, let us take $\lambda \sim 0.7$, which barely allows this model not to be ruled out by the cosmological data at the $95\%$ confidence level. Then, during the period from the beginning of the cosmological evolution to the present time the field $\phi$ changes by $\Delta \phi \sim 400$, which is a dramatic violation of the weak gravity conjecture advocated in \cite{Brennan:2017rbf}. 

A way to address this problem has been proposed in~\cite{Agrawal:2018own}. The authors suggest that in the early universe the potential is dominated by a term $e^{C(\phi) \phi}$ with $C(\phi) = \mathcal{O}(100)$. Then the field falls from $V=\mathcal{O}(1)$ to  $V\sim 10^{{-120}}$ within $\Delta \phi = \mathcal{O}(1)$, and it then enters the slow-roll quintessence regime with $\lambda < 1$. An example of a potential with the required properties would be
\be\label{100}
V = e^{\lambda\phi} + A\  e^{100 \phi}  \ ,
\ee
where $A$ is some constant. One might try to relate the large coefficient in the second exponent $10^{2}$ to $1/M_\text{GUT}$ \cite{Agrawal:2018own}, but it is not quite clear how this suggestion can be implemented.  As we have seen already, in the class of models considered in  \cite{Obied:2018sgi} it is very difficult to find an exponent coefficient $\lambda < \sqrt 2$. But it is equally difficult to find $\lambda \sim 100$. Indeed, in all the models that we were able to check, the exponent coefficients appeared as a result of  simple algebraic manipulations with numbers like $\sqrt{D-d}$, with $D = 10$ or $11$, and $d = 4$, so all of these exponent coefficients were of $\mathcal{O}(1)$.

Of course we may not need to know the dark energy potential all the way to the Planck density, but if we make a modest requirement that we want to know it at the nuclear density $\rho \sim 2\times 10^{14}$ g/cm$^{3}$, then the required excursion taking the potential $V\sim e^{\lambda\phi}$ with $\lambda < 0.7$ down to $V \sim 0.7 \times  10^{{-120}}$ would be $\Delta \phi > 140$. If we further require only that the potential is given by $V\sim e^{\lambda\phi}$ at a density smaller than the density of water,  then the required excursion would be $\Delta \phi > 90$. 

Finally, if we want to ensure the validity of the quintessence potential in the ultimately modest range of $10^{{-121}} < V_{0} \,   e^{\lambda\phi} < 10^{{-119}}$ for $\lambda = 0.7$, this would require that the theory should remain reliable in the super-Planckian range $\Delta \phi \sim 6.5$.  

Thus, according to the weak gravity conjecture, the potential $V\sim e^{\lambda\phi}$ can be used only in the immediate vicinity of the present value of the cosmological constant. Of course, the weak gravity conjecture is just a conjecture, but it is interesting that applying it to the string theory models of quintessence may lead to such an unexpected and somewhat disappointing conclusion.

\section{Discussion and conclusions}\label{disc}

Dark energy was  discovered 20 years ago~\cite{Riess:1998cb,Perlmutter:1998np}. This discovery created a turmoil  in theoretical physics, and in string theory in particular. At first, there was a hope that the discovery would go away, but this did not happen. The  first attempts to describe dark energy/quintessence in theories based on supersymmetry and supergravity were made in~\cite{Binetruy:1998rz,Brax:1999gp,Kallosh:2002gf}, and in M-theory in~\cite{Kallosh:2002gg,Gutperle:2003kc,Kallosh:2003bq}. The serious conceptual issues with quintessence in M-theory and supergravity were revealed soon, and no consistent string theory models of quintessence 
were found in $d = 10$ superstring theory. 

The situation changed with the invention of the 
KKLT scenario~\cite{Kachru:2003aw,Kachru:2003sx} and its various generalizations, such as KL~\cite{Kallosh:2004yh} and LVS~\cite{Balasubramanian:2005zx}, which suggested a multitude of possibilities to describe the present value of the cosmological constant in the context of the  string theory landscape~\cite{Douglas:2003um,Susskind:2003kw,Douglas:2006es,Denef:2007pq}. This theory is extremely complicated, and is far from being complete, but we have the proof of concept.  Due to the extreme multiplicity of the vacua, this scenario is very robust with respect to even very large quantum corrections, as illustrated in Fig.~\ref{chi}.

That is, of course, assuming that there are no no-go theorems proving that this whole set of ideas is internally inconsistent, and dS states are simply impossible in string theory. Here we stress that we are discussing no-go theorems, not the arguments in the spirit of the Wilsonian EFT, naturalness, weak or strong versions of the weak gravity conjecture, or the possibility that radiative corrections  strongly affect  dS vacua and reshuffle them  as shown in Fig. \ref{chi}. We are discussing real no-go theorems, which would state that all of the $10^{{500}}$ or more dS vacua in string theory cannot exist. Despite many attempts of many authors to prove such no-go theorems during the last 15 years, no such proofs are available.

In this paper, we have explained in detail that all the known no-go theorems of this kind can be evaded.
Here in this last section, we first summarize our statements concerning  the no-dS conjecture made in \cite{Obied:2018sgi}.
Their conjecture that dS vacua in string theory are not possible originates from their use of the original versions of the major no-go theorems. For the case of the Maldacena-Nunez no-go theorem~\cite{Maldacena:2000mw}, they assume nonsingular compactification manifolds. As  it is known for two decades,  such manifolds fail to describe chiral fermions in $d=4$~\cite{Acharya:2001gy}. Therefore, one should not require using nonsingular compactification manifolds for describing de Sitter geometries.  It is known that  one can evade the Maldacena-Nunez theorem taking into account higher-order curvature corrections and negative tension O-planes~\cite{Giddings:2001yu,Silverstein:2001xn}, as stressed in~\cite{Kachru:2003aw,Kachru:2003sx}.  
 
With regard to the no-go theorem of~\cite{Pilch:1984aw} on dS and supersymmetry, in pure dS supergravity one can evade the theorem by  involving a nonlinearly realized  supersymmetry, as it follows from the D-brane construction \cite{Aganagic:1996nn,Bergshoeff:2015tra,Hasegawa:2015bza}. It is this construction  that was used for the manifestly supersymmetric version of the KKLT construction in~\cite{Kallosh:2014wsa,Ferrara:2014kva,Bergshoeff:2015tra,Hasegawa:2015bza,Bergshoeff:2015jxa,Kallosh:2015nia,Kallosh:2016aep,Aalsma:2018pll}. Moreover, dS supergravity in $d=4$ is now derived  in the context of string theory compactification from $d=10$~\cite{Kallosh:2018nrk}.

The most recent criticism of the KKLT construction has been presented in~\cite{Sethi:2017phn} and~\cite{Moritz:2017xto,Moritz:2018ani}. A critical discussion of~\cite{Sethi:2017phn} can be found in~\cite{Kachru:2018aqn}.  A detailed analysis of~\cite{Moritz:2017xto,Moritz:2018ani} is given in~\cite{Kallosh:2018wme,Cicoli:2018kdo,Kallosh:2018psh}, as well as in Sections \ref{KKLT} and \ref{nogo} of this paper.  We believe that the 10d analysis of the KKLT scenario in~\cite{Moritz:2017xto} is unreliable \cite{Kallosh:2018wme,Cicoli:2018kdo}. Meanwhile, the 4d analysis performed in~\cite{Moritz:2017xto} was incorrect. According to~\cite{Kallosh:2018wme,Kallosh:2018psh}, all presently available consistent 4d generalizations of the KKLT construction, in the domain of their validity, confirm the existence of dS vacua in the KKLT scenario.

Of course, one can ignore this fact and simply speculate that stable or metastable dS vacua are impossible in a consistent quantum gravity theory, discard all models of dS space in a hope to make string theory great again, and then see what happens \cite{Brennan:2017rbf,Danielsson:2018ztv,Obied:2018sgi}. This no-dS conjecture takes us back to the situation we encountered 20 years ago, when we did not have any consistent description of the observational data in the context of string theory. This does not mean that any success in this direction is impossible, which is why we studied it, despite the fact that the motivation for the no-dS conjecture in  \cite{Brennan:2017rbf,Danielsson:2018ztv,Obied:2018sgi} does not seem convincing to us.

Returning to the discussion of the swampland, in~\cite{Agrawal:2018own} the analysis of quintessence is mixed with an early universe inflation. Here we stress that the basic no-dS conjecture is satisfied in all slow-roll inflationary models automatically. Indeed, the deviation from dS is the most important feature of all slow-roll inflationary models: the amplitude of inflationary perturbations blows up for $V_{,\phi} > V^{3/2}$. Therefore, there is no obvious reason to impose additional unmotivated constraints of the type of \rf{swamp} on these models. 

On the other hand, if one insists that the strong form of the no-dS conjecture \rf{swamp} should apply to inflationary models, it will be yet another argument against this conjecture. It is known from the latest Planck data release \cite{Akrami:2018odb} that $r \lesssim 0.064$ during the stage of inflation responsible for structure formation and CMB anisotropy in our part of the universe. This means that
\be
\epsilon = {1\over 2}\Big(  {V_{,\phi}\over V}\Big) ^2 = {1\over 2}c ^2 \lesssim 0.004 \, \implies c \lesssim  0.09\, .  
\ee
An analysis of related issues in~\cite{Agrawal:2018own,Kinney:2018nny}  gives similar constraints.  Therefore, if we would assume that the constraint  \rf{swamp} with $c \sim 1$ applies to inflation, we would conclude that the conjecture \rf{swamp} strongly contradicts the observational data.

In the case of dS versus quintessence for the late universe, things are more subtle and may need more attention since during the next decade major new cosmological data on the equation of state $w=p/\rho$ will be available. 
In this paper, we have asked the question `is the proposal~\rf{swamp}, which is consistent with string theory, compatible with the present data?'

We have performed a statistical analysis of the quintessence exponential potential $V_0 e^{\la \phi}$ with regard to the currently available cosmological data on the background expansion of the universe; see Section~\ref{observations}. In view of the controversy about the data  in~\cite{Agrawal:2018own,Heisenberg:2018yae} a due diligence was required. Namely, in~\cite{Agrawal:2018own} only a $2\sigma$ bound is proposed in the form $\la=c <0.6$ (in the first version of the paper $c<0.6$ was called a $3\sigma$ bound). In~\cite{Heisenberg:2018yae} it is suggested that their $1\sigma$ bound reproduces the $2\sigma$  bound in \cite{Agrawal:2018own}, while their $3\sigma$ bound is proposed in the form $c<1.35$. This way, the paper by Heisenberg et al.~\cite{Heisenberg:2018yae} has introduced an uncertainty which was necessary to resolve before we study string theory quintessence models. Our results, in the case of an analysis I, where we fixed the parameter $V_0$ to $0.7(3H_0^2)$ and varied the other parameters of the model, i.e. $\lambda$ and $\phi_0$, as well as $\Omega_\text{M}$, provided the one-dimensional, marginalized upper bounds on $c$,
\be\label{analysis1}
c \lesssim 0.13, \qquad c  \lesssim 0.54, \qquad  c\lesssim 0.87,
\ee
for $1\sigma$, $2\sigma$ and $3\sigma$ confidence levels, respectively; see Fig.~\ref{fig:mainConstraints_V0p7}.
In an analysis II, we performed a statistical exploration of the parameter space when $V_0$ was allowed to vary as well, in order to take care of some statistical subtleties around $\lambda=0$. In that case, depending on the prior on $\phi_0$, i.e. the range over which $\phi_0$ was allowed to vary, the upper bounds on $c$ varied: the broader the range of $\phi_0$, the lower the upper bounds on $c$. We fixed $\phi_0$ to zero (i.e. we set the $\phi_0$ range to zero) in order to obtain the largest upper bounds on $c$. The marginalized, one-dimensional upper bounds on $c$ in this case are
 \be\label{analysis2}
c\lesssim0.49, \qquad c\lesssim0.80, \qquad  c\lesssim1.02,
\ee
for $1\sigma$, $2\sigma$ and $3\sigma$ confidence levels, respectively; see Fig.~\ref{fig:mainConstraints_phi00}. An additional profile likelihood analysis of the parameter space in a frequentist approach (as opposed to our main Bayesian marginal posterior analysis) provided bounds very close to~(\ref{analysis2}). This demonstrated that (\ref{analysis2}) were indeed the weakest and most prior-independent bounds on $\lambda$ that one could obtain with the cosmological data used in this paper. By including additional information beyond the purely geometrical tests of the background expansion as we have employed in our paper, e.g. by adding the perturbative information from the Planck CMB temperature and polarization data, we expect these bounds to be even tighter.
Thus, if we would like to be as tolerant to theoretical cosmological models as possible with regard to the data, we can say that models with $c \lesssim 1.02$ can be looked at more carefully since they are not immediately ruled out by the data. But, one should keep in mind that this is a last resort, as all models with $c \gtrsim 1.02$ should be dismissed without hesitation.

On the theoretical side, we went through the list in~\cite{Obied:2018sgi} with an update provided in~\cite{Roupec:2018fsp} to make sure that the models in $d=4$ which we confront with the data were viewed as string constructions, beyond speculations. In particular, we have noticed that all the models with 
$
c<\sqrt 2
$
suggested in~\cite{Obied:2018sgi} do not really belong to string theory or M-theory constructions. Meanwhile, the data can only tolerate $c\lesssim 1.02$ at most, as we have explained in Section~\ref{observations}, and therefore the models that the authors of~\cite{Obied:2018sgi} are left with, requiring $c\gtrsim1.4$, are all ruled out. The additional quintessence models added in  \cite{Agrawal:2018own}  are either ruled out by the data, or represent inflationary models with moduli stabilization converted to models of quintessence. All of them contradict the condition \rf{swamp}. 

But this is not the only problem with the string theory models of quintessence. There are many general conceptual issues discussed in Section~\ref{problems} that must be addressed: decompactification, fifth force, quantum corrections, consistent embedding of dark matter, the problem of initial conditions, etc. On top of that, our analysis brings some surprising news to those who believe in the weak gravity conjecture: The potential $V\sim e^{\lambda\phi}$, or any other similar potential which can be used for describing quintessence in terms of a canonically normalized scalar field $\phi$, is well defined only in the tiny range of values of the potential in the immediate vicinity of the present value of the cosmological constant. 

This suggests that if a consistent theory of quintessence can be constructed in the context of string theory, it will not replace the string theory landscape scenario, but will rather enhance it, by adding to the many dS minima, which are shown in Fig.~\ref{chi}, a collection of  segments of the potential, which should be very flat, with ${| \nabla_\phi V|\over V} < 1$.  Note that the last requirement is directly opposite to the  strong form of the  no-dS conjecture \rf{swamp}. 

In preparation for potential deviations of the dark energy data from a cosmological constant provided by upcoming or future surveys, we have already constructed quintessential inflation models in string theory motivated versions of supergravity~\cite{Akrami:2017cir},  which might fit such future data. However, a majority of supergravity models compatible with the early universe inflation end up in dS minima and explain dark energy via a cosmological constant taking  values in the landscape. 

 Recent cosmological observations have attracted attention to specific ideas/aspects of non-perturbative superstring theory, which are helpful in building models compatible with the data. 
The $\overline {D3}$-brane is a source of nonlinearly realized supersymmetry and positive contribution to energy, which is a characteristic property of the KKLT uplifting procedure. In phenomenological model building, the corresponding nilpotent multiplet, in addition to providing positive energy, plays the role of a stabilizer superfield, which allows an advanced version of $\alpha$-attractor models \cite{Kallosh:2013yoa}; we called them geometric inflation~\cite{Kallosh:2017wnt,McDonough:2016der}. These models provide a good fit to the Planck 2018 data~\cite{Akrami:2018odb}; see Fig.~\ref{Precision}.

Thus,    the ideas originating in string theory with dS vacua   influenced the construction of phenomenological $d=4$ supergravity models  of inflation  compatible with the cosmological data. These are the corners of string theory where better understanding and more developments may be useful since they are already in the sweet spot of the data, i.e. in the blue area of the $r-n_s$ plane in Fig.~\ref{Precision}.

\

\noindent{\bf {Acknowledgments:}}  We are grateful to  E. Bergshoeff,  S. Ferrara, D. Freedman and A. Van Proeyen for the clarification of the conceptual issues of de Sitter supergravity and of the no-go theorem  \cite{Pilch:1984aw}. 
We are also grateful to the organizers and participants of the conference `Dark Side of the Universe' in Guadeloupe, June 25-29, 2018, especially J. Garcia-Bellido  and T. Eifler, for stimulating discussions on the dark energy data. We had  very useful discussions of string theory, cosmology and supergravity with S. Kachru, E. McDonough, M. Scalisi, E. Silverstein, R. Wechsler, A. Westphal, T. Wrase, Y. Yamada and M. Zagermann. We also thank O. Contigiani, M. Martinelli and A. A. Sen for helpful discussions on the statistical analysis performed in this paper. Y.A. acknowledges support from the Netherlands Organization for Scientific Research (NWO) and the Dutch Ministry of Education, Culture and Science (OCW), and also from the D-ITP consortium, a program of the NWO that is funded by the OCW. The work  of R.K. and A.L. is supported by SITP,   by the NSF Grant PHY-1720397, and by the Simons Foundation grant. V.V. is supported by a de Sitter PhD fellowship of NWO.

\appendix

\section{What was the problem with de Sitter supergravity in the past?}\label{dSsupergravity}

The no-go theorem for dS vacua in supergravity \cite{Pilch:1984aw} was proposed and proven in 1985 in the context of pure supergravity, i.e. in supergravity without matter multiplets.  For quite a while, this theorem was considered to be a real obstacle preventing dS vacua in supergravity. Later on, with the development of phenomenological supergravity including chiral supermultiplets with scalar fields $Z$ as their first component, the situation changed dramatically.

The F-term part of the scalar field potential in supergravity can be represented by the well-known expression 
\be 
V(Z^i, \bar Z^{\bar i})=e^{K}\Big(|D_iW|^{2}-3|W|^{2}\Big)\ ,
\label{2d}
\ee
where $W(Z)$ is the superpotential, $K(Z,\bar Z)$ is the \K\ potential, and supersymmetry breaking is controlled by the value of $ D_i W$.
As the simplest example, consider the model with the minimal {K\"ahler} potential $K = Z \bar Z$ and the linear superpotential $W(Z) = m^2(Z+\beta)$. This is the famous Pol\'{o}nyi model \cite{Polonyi:1977pj} of 1977, described in most textbooks on supergravity.
One can represent the complex field $Z$ as a sum of two canonically normalized fields $\phi$ and $\chi$, $Z = (\phi+i\chi)/\sqrt 2$. One can show that the minimum of the effective potential occurs at $\chi = 0$, so one can restrict themselves to the investigation of $V(\phi)$, where
\be
 V(\phi) = {m^4 }\ e^{\phi^2\over 2 }\Bigl[\Bigl(1+{\phi\over \sqrt2  }\Bigl({\phi\over \sqrt2  } +\beta\Bigr)\Bigr)^2  -    3\Bigl({\phi\over \sqrt2  }+\beta\Bigr)^2\Bigr].
\label{Pol} \ee
For $\beta = 2-\sqrt 3$, this potential has a stable minimum at $V = 0$, but if one slightly decreases $\beta$, this minimum becomes a stable dS minimum at $V>0$; see Fig. 8 in~\cite{Kallosh:2002gf}.

Thus, despite the no-go theorem  \cite{Pilch:1984aw}, one can easily find dS vacua in realistic supergravity models including matter fields. Why was this theorem then discussed in~\cite{Brennan:2017rbf,Danielsson:2018ztv} as a part of the general no-dS argument, and what is the  way to evade this theorem altogether~\cite{Kallosh:2014wsa,Ferrara:2014kva,Bergshoeff:2015jxa,Bergshoeff:2015tra,Hasegawa:2015bza,Kallosh:2015nia,Kallosh:2016aep,Aalsma:2018pll}? 

The mindset of the string theory community about supersymmetry and de Sitter space can be traced back to the lectures on dS/CFT correspondence by Strominger et al.~\cite{Spradlin:2001pw}. An interpretation of the results of~\cite{Pilch:1984aw} was proposed there in an attempt to formulate dS/CFT correspondence. It was explained in~\cite{Spradlin:2001pw}, with reference to~\cite{Pilch:1984aw}, that de Sitter space was inconsistent with the existence of the supergroup that includes isometries of de Sitter space and has unitary representations. That helped to explain why the CFT in dS/CFT had non-unitary representations. 
Indeed, in  extended $\cN> 1$ supersymmetry in $d=4$ the superalgebra with $\cN$ generators $Q_\alpha ^i$ is available, it has  no problem with Jacobi identities differently from the $\cN=1$ case, but the  representations of the superalgebra  are non-unitary.

{\it Meanwhile, in  $\cN=1$ there are no supergroups that would include isometries of de Sitter space. This fact has nothing to do with non-unitarity}.
Let us look more carefully at  the no-go theorem of~\cite{Pilch:1984aw} for the case of $d=4$ and $\cN=1$, which is interesting for observational cosmology.
The bosonic algebra for 
 either de Sitter $SO(4,1)$ space or anti-de Sitter  $SO(3,2)$ space is
\be
[P_\mu , P_\nu]= \pm {1\over 4L^2} M_{\mu\nu} \ .
\label{P}\ee
Here, the upper sign is for dS and the lower sign is for AdS. To include a supersymmetry generator $Q$, one has to add additional elements to make it a superalgebra, 
\be
[P_\mu, Q_\alpha] = {1\over 4L} (\gamma_\mu Q)_\alpha \ , \qquad  \{Q_\alpha, Q_\beta\}= -{1\over 2} \gamma^\mu_{\alpha \beta} P_\mu - {1\over 8L}\gamma^{\mu\nu}_{\alpha \beta} M_{\mu\nu} \ .
\label{PQ}\ee
 {\it One finds,   using Jacobi identities, that the $\cN= 1$ superalgebra in \rf{P}-\rf{PQ}  is consistent only for the lower sign in \rf{P}, i.e. for the AdS case}.   
This explains why pure AdS supergravity (supergravity without matter multiplets) with a negative cosmological constant and unbroken supersymmetry is known for 4 decades~\cite{Townsend:1977qa}.  Meanwhile, pure dS  supergravity with a positive cosmological constant was constructed only in 2015. 

When pure $\cN=1$ de Sitter supergravity was constructed in 
 \cite{Bergshoeff:2015tra,Hasegawa:2015bza}, it was important to clarify how it would evade the no-go theorem of  \cite{Pilch:1984aw}, and why  dS vacua could exist even in the absence of any scalar fields. 
This happens because the no-go theorem \cite{Pilch:1984aw} is valid only for theories where there is a supersymmetry generator which flips a one-bosonic state into a one-fermionic one, and vice versa. Meanwhile, in the Volkov-Akulov nonlinear  supersymmetry case, the goldstino multiplet has only a fermion state. The partner of a single fermion state is a two-fermion state, and supersymmetry is spontaneously broken. Physical states in such a theory are not representations of the superalgebra, which is  why dS supergravity with a positive cosmological constant evades the no-go theorem of \cite{Pilch:1984aw}.

This is very much like in the Coleman-Mandula  case. It is true that space-time and internal symmetries cannot be combined in any but a trivial way in theories where only Poincar\'e algebra defines the representations of the states in the theory. However, if one adds supersymmetry generators, a new theory is created, where space-time and internal symmetries are combined in a nontrivial way. 
      
Linearly realized $d=4$ global supersymmetry was discovered in 1971-1974~\cite{Golfand:1971iw,Wess:1973kz}. A local, linearly realized supersymmetry (supergravity) was discovered in 1976 \cite{Freedman:1976xh,Deser:1976eh}.  AdS vacua in pure supergravity were discovered in 1977~\cite{Townsend:1977qa}, one year after the discovery of supergravity with a vanishing cosmological constant.

Nonlinearly realized supersymmetry was discovered by Volkov and Akulov in 1972~\cite{Volkov:1972jx}, and then in D-brane physics by John Schwarz et al. in 1997 \cite{Aganagic:1996nn, Kallosh:1997aw}, and finally its local version,  pure supergravity with dS vacua,  was constructed  in \cite{Bergshoeff:2015tra,Hasegawa:2015bza}, 43 years after   \cite{Volkov:1972jx}.

This result plays a crucial role in the advanced version of the KKLT construction  \cite{Kallosh:2014wsa,Ferrara:2014kva,Bergshoeff:2015jxa,Bergshoeff:2015tra,Hasegawa:2015bza,Kallosh:2015nia,Kallosh:2016aep,Aalsma:2018pll}. Importantly, it applies only to the models with a single $\overline {D3}$-brane, which was the case in the original KKLT construction \cite{Kachru:2003aw,Kachru:2003sx}. The local fermionic $\kappa$-symmetry of a single $\overline {D3}$-brane \cite{Aganagic:1996nn}  leads to a nonlinear realization of supersymmetry. When a nonlinear realization of supersymmetry originating from a single $\overline {D3}$-brane is taken into account, see \cite{Aganagic:1996nn,Kallosh:1997aw,Bergshoeff:2013pia}, and \cite{Kallosh:2014wsa,Bergshoeff:2015jxa} for details, one finds a universal source of the positive energy of space-time,
\be
{\cal L }_{\overline {D3}}= -T^3 \int d^4 \sigma \sqrt {- \det G_{\mu\nu}} = -T^3 \int \det E =  -T^3 \int E^0\wedge E^1\wedge E^2 \wedge E^3 = {\cal L }_\text{VA} \ ,
\ee
where the supersymmetric 1-form $E^a$ depends on the Volkov-Akulov fermion goldstino $\theta(\sigma)$,
\be
E^a= d\sigma^\mu e_\mu^a (\theta (\sigma)) = d\sigma^a+ \theta \Gamma^a d\theta \ .
\ee
The positive vacuum energy at $\theta=0$ is
\be
H_{\overline {D3}}|_{\theta=0}= - {\cal L }_{\overline {D3}}|_{\theta=0} = T^3 \int \det E |_{\theta=0}  = T^3 >0 \ .
\ee

The nilpotent multiplet representing an $\overline {D3}$-brane can play an important  role not only in the KKLT uplifting, but also in the effective supergravity models describing inflation. The nilpotent multiplet facilitates the stabilization of other moduli, leaving only one light inflaton. The geometric inflation models proposed in  \cite{Kallosh:2017wnt}  can be interpreted as associated with the geometry of the $\overline {D3}$-brane in the moduli space background. The geometry of the nilpotent multiplet $G_{S\bar S}$ carries all the information about inflation in these models. This framework helps to develop broad classes of $\alpha$-attractor models in~\cite{Kallosh:2013yoa}, which fit the data very well; see Fig.~\ref{Precision} describing the status of various inflationary models after the Planck 2018 data release~\cite{Akrami:2017cir}.

The effective $d=4$ dS supergravity, which includes interaction with any number of chiral  multiplets, in addition to a nilpotent multiplet with $S^2(x,\theta) = 0$, has been described in~\cite{Kallosh:2015sea,Kallosh:2015tea,Schillo:2015ssx,Kallosh:2016ndd,
Ferrara:2016een,DallAgata:2016syy,Freedman:2017obq}. The potential in these models is
\be 
V(Z^i, \bar Z^{\bar i}; S, \bar S)=e^{K}\Big(F^2+ |D_iW|^{2}-3|W|^{2}\Big)\ ,
\label{nilp}
\ee
where $F$  is a necessarily non-vanishing value of the auxiliary field of the nilpotent multiplet.
The chiral nilpotent multiplet has components $\Big ({\chi^2\over 2 F} ,   \chi ,  F \Big )$,  where $\chi$ is the fermionic goldstino and the first component is a bilinear in goldstino. There are no scalars, and the consistency  of this scalar-less multiplet requires $F\neq0$.
De Sitter vacua  at $V_{,\phi}=0$ are present under the condition $|F^ 2| -3|W |^2 > 0$. This explains the origin of eq.~\rf{dif} with $V_{\overline {D3}}=e^{K}F^2$.

It is instructive to compare this mechanism to the dS uplifting in the Pol\'{o}nyi model \rf{Pol}. This model is very simple, and it was routinely used in supergravity phenomenology for nearly 4 decades. However, it was  difficult to find a  string theory interpretation of the Pol\'{o}nyi field. In dS supergravity, uplifting is achieved due to the nilpotent multiplet, without introducing extra scalars. This mechanism does have a string theory interpretation in terms of the $\overline {D3}$-brane contribution to the energy density in the KKLT construction~\cite{Kallosh:2014wsa,Ferrara:2014kva,Bergshoeff:2015jxa,Bergshoeff:2015tra,Hasegawa:2015bza,Kallosh:2015nia,Kallosh:2016aep,Aalsma:2018pll}.

\section{Accelerating universe according to  supercritical string theory}\label{supercritical}

One may wonder whether the negative conclusions of our investigation in Section~\ref{models} are generic or there are exceptions. In particular, is it possible to modify a theory and find models with small $c$?

The models of accelerating universe based on supercritical bosonic string theory have been studied in~\cite{Dodelson:2013iba} for the space $\mathbf{ R}^{d-1,1} \times \mathbf{ T}^n$.
To compare with the data, we consider only compactifications to $d=4$.  The total number of dimensions is 
$
D= 4+n \gg 26
$.
This number is supposed to be  much greater than the critical dimension of the bosonic string theory, $D_\text{cr}=26$.
There are three classes of highly elaborated string theory models derived in~\cite{Dodelson:2013iba}. In the limit of small $V$ (large $\phi$), the potentials are reduced, after the stabilization of extra moduli, to a single exponential for a canonical field such that 
\be 
V_\text{DE}\sim e^{-\lambda \phi}\, ,  \qquad \lambda^2= {2\over K} \ .
\ee
It has been explained in~\cite{Dodelson:2013iba} that the models of accelerating universe derived in string theory may still be subject to Standard Model corrections, and should not be viewed as realistic. We use them here just for illustrative purposes of comparison with the improved data on dark energy, since the Standard Model corrections are not easy to evaluate.

The first class of models given in eq.~(2.42) of~\cite{Dodelson:2013iba} has
the minimal value $\lambda \simeq 1.2$. Another class of models given in eqs.~(3.9) and (3.10) of~\cite{Dodelson:2013iba} requires 
$
\lambda \approx 1.34
$. Both models are ruled out by the data.
Finally, a class of models described by eq.~(2.57) in~\cite{Dodelson:2013iba} gives
\be
K= 4k+ {1\over 64k^2 - 28 k -1} \ ,
\ee
where $k$ is an arbitrary even integer, $k\geq 2$. Here $\lambda^2\sim  {1\over 2k} $, and therefore, one can have
\be
\la \lesssim  0.5 \ .
\ee
With this value of $\lambda$, the model can be used for quintessence. However, this requires an internal space with dimension
\be
n=D-4= (64k^2 - 28 k -1) (32k-6) \geq 11542 \ .
\ee
Potential conceptual issues related to quantum corrections, possible light KK modes, and growth of volume of the compactified multi-dimensional space in the supercritical  bosonic string theory have to be studied.  

\section{Comparison to previous bounds on $\lambda$}\label{sec:priorbounds}

In this appendix, we compare the bounds we found on $\lambda$ in Section~\ref{observations} with the ones presented in the recent literature, i.e. in~\cite{Agrawal:2018own,Heisenberg:2018yae}. The authors of these papers have not performed a detailed statistical analysis of the exponential model $V_0 e^{\lambda\phi}$, and have based their studies on a simple inference based on the constraints provided in~\cite{Scolnic:2017caz} on the Chevallier-Polarski-Linder (CPL) parametrization~\cite{2001IJMPD..10..213C,2003PhRvL..90i1301L} of dark energy. The CPL parameterization approximates the dark energy equation of state $w_\text{DE}$ by
\begin{equation}
w_\text{DE}(z)= w_0 + w_az/(1+z),\label{eq:CPL}
\end{equation}
where $w_{0}$ and $w_{a}$ are constants and $z$ is the redshift. This parametrization is a valid approximation to the evolution of dark energy equation of state for an arbitrary dark energy model as long as one stays close to $z=0$, i.e. to the present time. Before we compare our results with the results of~\cite{Agrawal:2018own,Heisenberg:2018yae}, let us try to follow the recipe given in those papers and reproduce the bounds provided there. This is important and instructive, and shows how one may obtain different results without performing a rigorous statistical analysis of a cosmological model. This also explains the differences between our methodology and the ones used in~\cite{Agrawal:2018own,Heisenberg:2018yae}.

\begin{figure}[t!]
\begin{center}
\includegraphics[scale=0.33]{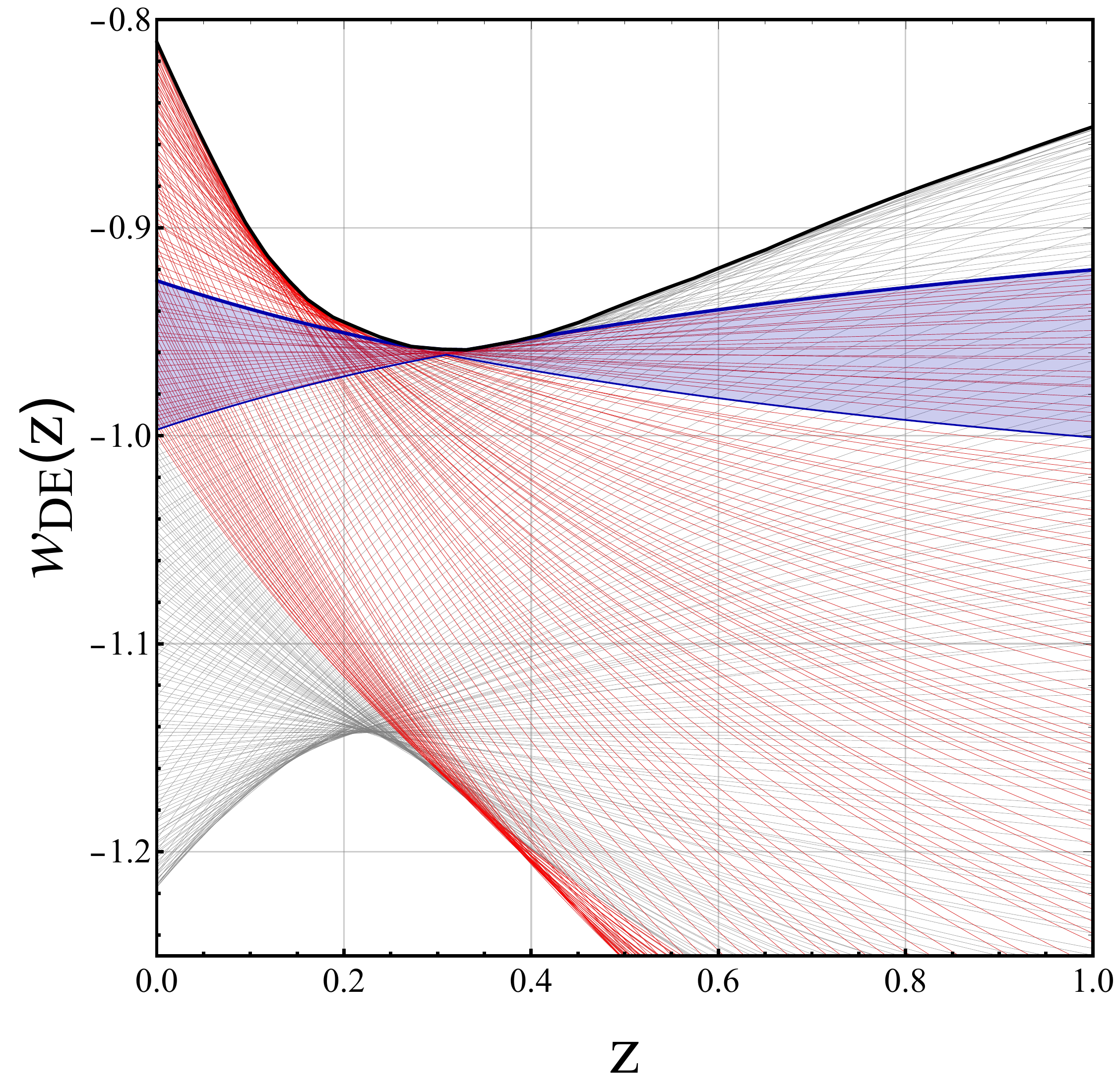}
\includegraphics[scale=0.33]{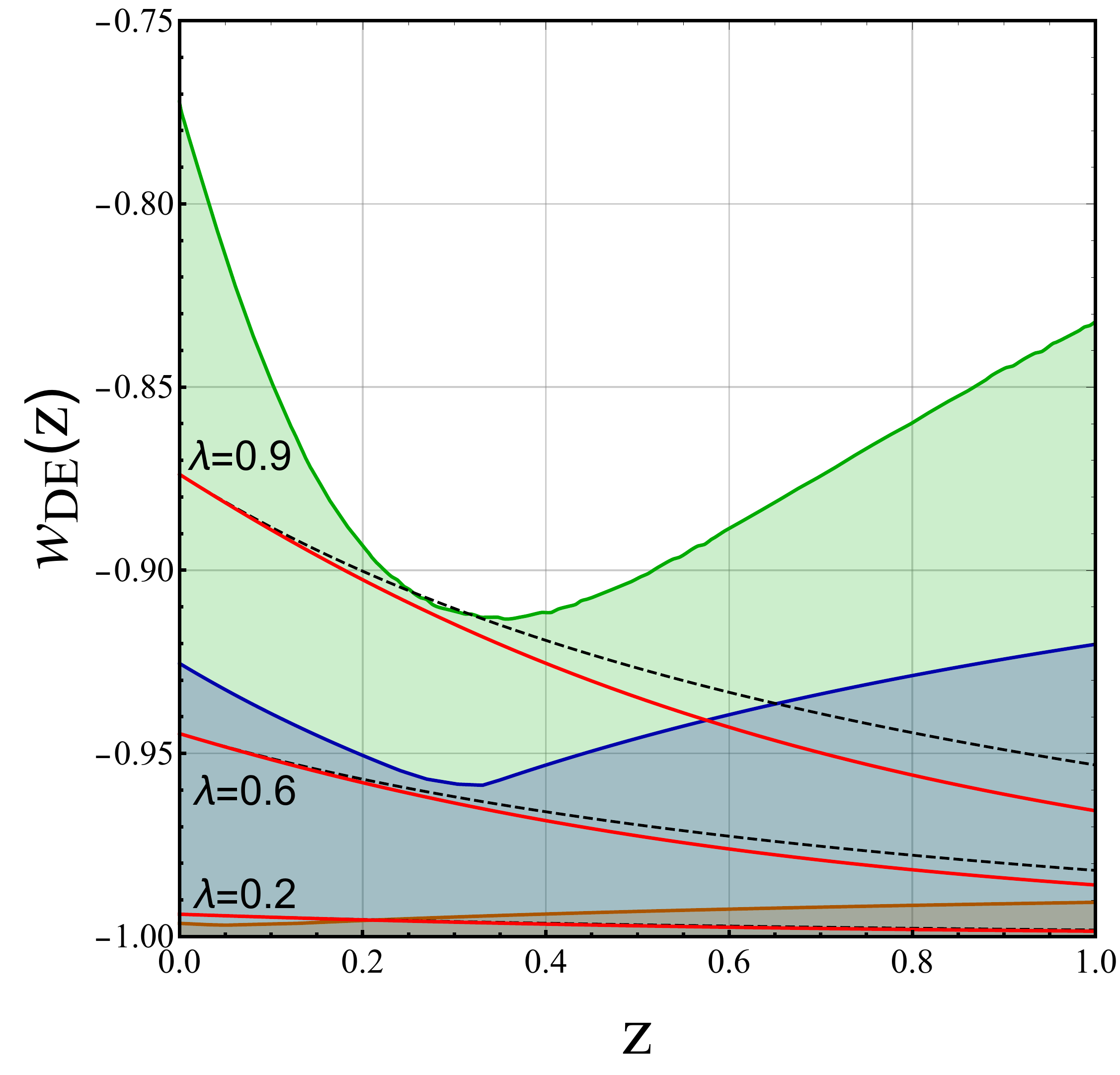}
\end{center}
\caption{\footnotesize {\it Left panel:} Reconstruction of the bound on the evolution of dark energy equation of state $w_\text{DE}$ provided by Agrawal et al. in Fig.~1 of~\cite{Agrawal:2018own}. The gray curves depict viable CPL-based $w_\text{DE}(z)$ corresponding to the $95\%$ contour in the $w_0-w_a$ plane given in Fig.~21 of~\cite{Scolnic:2017caz}. The upper envelope of these curves (thick, black curve) agrees with the exclusion curve provided in Fig.~1 of~\cite{Agrawal:2018own}. The blue curves show solely the curves which do not cross the phantom divide at any redshift $z<1$. {\it Right panel:} The thick, blue curve shows the $95\%$ exclusion bound on $w_\text{DE}$ obtained through the procedure of the left panel (corresponding to the upper envelope of blue curves). The orange curve shows the predicted exclusion bound inferred from the forecast analysis of~\cite{Sprenger:2018tdb} for the upcoming large-scale structure surveys Euclid and the SKA, in combination with the CMB constraints from Planck. The red curves show the exact, numerically calculated $w_\text{DE}(z)$ for three values of $\lambda=0.2, 0.6, 0.9$, while the dashed, black curves depict the corresponding CPL approximation according to eq. (\ref{eq:CPL}). For comparison, we have also shown (in green) the $95\%$ exclusion curve provided by Heisenberg et al. in~\cite{Heisenberg:2018yae}. See the text for a discussion of why this curve does not agree with the blue curve, although the analysis of~\cite{Heisenberg:2018yae} is a Fisher-matrix based approximation to the procedure illustrated in the left panel of this figure.}
\label{fig:CPL}
\end{figure}

We start with digitizing the perimeter of the $95\%$ confidence region provided in Fig.~21 of~\cite{Scolnic:2017caz} (the outer yellow contour), which contains all the {\it viable} combinations of the CPL parameters $w_0$ and $w_a$. This contour has been obtained by combining the constraints from the Pantheon supernovae data with the ones from the CMB, BAO and the local measurements of $H_0$. We then plot $w_\text{DE}(z)$ for all the sampled points according to eq. (\ref{eq:CPL}). This provides us with the set of gray curves shown in the left panel of Fig.~\ref{fig:CPL}. Comparing these curves to the bound provided in Fig. 1 of~\cite{Agrawal:2018own}, we immediately realize that the upper envelope of our curves (depicted by a thick, black curve) agrees perfectly with what is referred to by Agrawal et al. in~\cite{Agrawal:2018own} as the $2\sigma$ bound.

This bound, however, includes dark energy evolutions which become phantom (i.e. with $w_\text{DE}(z)<-1$) for some periods during the history of the universe. Since quintessence models like our exponential potential cannot produce phantom behavior and do not cross the phantom divide, we can remove all the curves which cross $w_\text{DE}=-1$ at some point. Excluding first all the cases with $w_0<-1$, we are left with the red curves in the figure. Restricting ourselves next to $z<1$ (where the CPL approximation holds), by applying the no-phantom condition to all times we can further remove some of the red curves; this yields the blue curves. By comparing the upper envelope of the remaining blue curves (the thick, blue curve) with that of the original gray curves, we see that our $95\%$ exclusion curve is now different from the bound presented by Agrawal et al. in~\cite{Agrawal:2018own}, where the phantom solutions have not been removed. This is however not important for imposing bounds on $\lambda$ through this procedure, since the exclusion comes solely from the valley of the envelope, which is the same independently of including or excluding the phantom evolutions. By comparing the solutions for $w_\text{DE}$ corresponding to different values of $\lambda$, we see that the $w_\text{DE}$ curve for $\lambda\approx 0.65$ touches the bottom of the $95\%$ exclusion curve, and therefore one may infer the upper bound of $\lambda < 0.65$ with $95\%$ confidence.

Even though this {\it inferred} bound fully agrees with the finding of~\cite{Agrawal:2018own}, we again emphasize that one is required to perform a consistent statistical analysis of the model, as we have done in Section~\ref{observations}, in order to correctly constrain $\lambda$. It is true that any viable point in the parameter space must produce a $w_\text{DE}$ curve below the bound found through the procedure of Fig.~\ref{fig:CPL} (assuming that the CPL parametrization is a good approximation to the exact solution), but the opposite may not be correct. The curves in Fig.~\ref{fig:CPL} capture only the properties of the dark energy equation of state, and they do not contain information on the density of dark energy, which is crucial for the viability of the model. Particularly, these curves have been obtained by fixing the value of $V_0$ (and $\phi_0$), as well as $\Omega_\text{M}$, and they do not include the observational uncertainties on these quantities. 

In the right panel of Fig.~\ref{fig:CPL}, we show three exclusion curves derived based on the CPL approximation described above. The blue curve is identical to the $95\%$ exclusion curve obtained in the left panel of the figure, which is in agreement with the bound found by Agrawal et al. in~\cite{Agrawal:2018own}. We have shown the evolution curve corresponding to $\lambda=0.6$ for comparison. The solid, red curve shows the numerically calculated $w_\text{DE}(z)$, and the dashed, black curve close to it shows the corresponding CPL approximation according to eq. (\ref{eq:CPL}). This demonstrates that the CPL approximation, although not being identical to the exact solution, is not very far from it. The exclusion plot and the value of $\lambda$ agree with Fig. 1 of~\cite{Agrawal:2018own}.

We have also shown in the same plot the exclusion curve presented by Heisenberg~et~al. in~\cite{Heisenberg:2018yae} as a $2\sigma$ bound (green exclusion curve in the right panel of Fig.~\ref{fig:CPL}), as well as the evolution curve corresponding to $\lambda=0.9$ reported in~\cite{Heisenberg:2018yae} as the $2\sigma$ upper bound on $\lambda$. Even though the authors of~\cite{Heisenberg:2018yae} seem to have found this curve through an analysis of the same $95\%$ contour of~\cite{Scolnic:2017caz} (the outer yellow contour in Fig. 21), their exclusion curve is clearly very different from what we have found from the exact same contour (the blue curve in Fig.~\ref{fig:CPL}). Although the CPL-based method presented in~\cite{Heisenberg:2018yae} is slightly different from our CPL-based procedure here, and is based on a Fisher matrix and elliptical approximation to the confidence contours, we do not expect it to give significantly different bounds if the ellipses well approximate the actual contours. We tried to reproduce the results of~\cite{Heisenberg:2018yae} by repeating the same procedure as described by the authors, but failed to obtain their curve. This gave us an exclusion curve which resembled more the blue curve in the right panel of Fig.~\ref{fig:CPL} than the curve presented in~\cite{Heisenberg:2018yae}, as expected. The authors of~\cite{Heisenberg:2018yae} also report a $99.7\%$ ($3\sigma$) upper bound of $1.35$ on $\lambda$, which is in strong disagreement with our statistical results, as the largest $3\sigma$ upper bound we found for $\lambda$ through our rigorous statistical analysis was $\lambda \sim 1.02$, which is robust and reliable.\footnote{It is important to note that the data sets used in our analysis did not include all the information used in~\cite{Scolnic:2017caz} to obtain their confidence contours on dark energy (adopted by Heisenberg et al. in~\cite{Heisenberg:2018yae}). We therefore expect the bound on $\lambda$ to become even tighter than $\sim 1$ when the observational information applied in~\cite{Scolnic:2017caz} is fully used. Therefore, the disagreement between our results and the approximate ones provided in~\cite{Heisenberg:2018yae} is expected to become even stronger if we use the same data sets as those used in the analysis of~\cite{Heisenberg:2018yae}.}

Finally, we have shown in the right panel of Fig.~\ref{fig:CPL} also an exclusion curve obtained through the same CPL procedure described above, but for the upcoming Stage IV large-scale structure surveys (orange curve). This curve is based on the $95\%$ forecast contour constructed in~\cite{Sprenger:2018tdb} for the combination of Euclid~\cite{Laureijs:2011gra} and the Square Kilometre Array (SKA)~\cite{SKARedBook2018} data sets, where the constraints from Planck~\cite{Akrami:2018vks} are also added. This exclusion curve shows that the upcoming cosmological surveys should be able to constrain $\lambda$ to values as low as $\sim 0.15$ (with $95\%$ confidence). We, however, emphasize that this bound is based on a simple CPL-based forecast, which should be considered only as a rough estimate. Our intention here has solely been to verify the similar CPL-based forecast of~\cite{Heisenberg:2018yae}. Our results seem to imply that by combining the information from several Stage IV surveys, one can constrain $\lambda$ more strongly than is predicted in~\cite{Heisenberg:2018yae}, which is based on Euclid alone. Nevertheless, in order to correctly estimate the future bounds one is required to perform a proper and detailed forecast analysis, which is beyond the scope of the present work. 


\paragraph*{\textbf{Note added after the appearance of~\cite{Heisenberg:2018rdu}:}} In a recent note~\cite{Heisenberg:2018rdu}, which appeared after an earlier version of the present paper, the authors of~\cite{Heisenberg:2018yae} have commented on our discussion above, and on the differences between the results of their analysis and those we have provided in the present paper.

Let us first mention that even if one takes the value quoted by~\cite{Heisenberg:2018yae} as the {\it current} $3\sigma$ bound on the parameter $\lambda$, our general conclusions in this paper will not be altered. Indeed, as we have showen in Section~\ref{models}, all string theory examples corresponding to the specific string theory constructions of~\cite{Obied:2018sgi} require $\lambda=c\gtrsim \sqrt 2 $, and all of them are therefore ruled out by the cosmological data with more than $99.7\%$ confidence. Therefore, the difference between $1$ and $1.35$ does not change this main result. Nevertheless, it is important to understand why the constraints on $\lambda$ obtained by Heisenberg et al.~\cite{Heisenberg:2018yae} are much weaker than the constraints obtained by Agrawal et al. in~\cite{Agrawal:2018own} (confirmed in Fig.~\ref{fig:CPL}) and the ones found in Section~\ref{observations} of the present work. 

Our approach in this work has been based on the standard Bayesian parameter estimation framework and a rigorous MCMC data analysis (verified by an independent, frequentist, profile likelihood analysis), which is widely employed in cosmology, while the analysis of~\cite{Heisenberg:2018yae} is based on approximating the existing constraints on the CPL parameters by a multivariate Gaussian distribution, which is not suitable for performing a reliable parameter estimation. It is clear that our approach does not suffer from the same methodological issues that exist in the type of analysis presented in~\cite{Heisenberg:2018yae}. Therefore, we cannot agree with the authors of~\cite{Heisenberg:2018rdu} in that their results, contradicting our results and the results of~\cite{Agrawal:2018own}, are ``more realistic.''    

From a technical point of view, we would like to mention again that if one approximates the $1\sigma$ contour of~\cite{Scolnic:2017caz} by an ellipse centered at the best-fit point given in~\cite{Scolnic:2017caz}, adjusts the principal components and the orientation of the ellipse with those of the actual contour, and repeats the procedure as described in~\cite{Heisenberg:2018yae}, then one obtains a $2\sigma$ curve which resembles the blue curve in the right panel of Fig.~\ref{fig:CPL} (though, of course, not matching exactly, because it is still only an approximation to the actual CPL contour), which is in agreement with~\cite{Agrawal:2018own} but not with~\cite{Heisenberg:2018yae}. 

The authors of~\cite{Heisenberg:2018yae} have not provided information about the exact centers of the ellipses and their widths and orientations, and for that reason, it was not clear to us why their procedure gave results in disagreement with our Fig.~\ref{fig:CPL} and the results of~\cite{Agrawal:2018own}. However, in their recent note~\cite{Heisenberg:2018rdu}, the authors state that their ``bounds on $\lambda$ are very sensitive to the precise values of the best-fitting parameters as well as on the principal values and the orientation of the principal axes of the inverse Fisher matrix.'' Based on this statement, and in an attempt to understand the source of the mismatch between their curve and those in Fig.~\ref{fig:CPL} and in~\cite{Agrawal:2018own}, we noticed that by particularly centering the ellipse at the point with $w_0=-1$ and $w_{a}=0$ (corresponding to $\Lambda$CDM), one could reproduce the $2\sigma$ bound of~\cite{Heisenberg:2018yae} quite closely. Let us state, however, that such an arbitrary choice of the center is incorrect and unjustified. In the process of parameter estimation, we should use the real data, not our expectations; in our MCMC data analysis, this is done automatically. In the approximate data analysis performed in~\cite{Heisenberg:2018yae}, one should place the centers of the ellipses in such a way that they approximate the real data distribution given in~\cite{Scolnic:2017caz}, for which the centers differ from $w_0=-1$ and $w_{a}=0$.

In summary, if we use the approximate method of~\cite{Heisenberg:2018yae} properly, we can closely reproduce the results of Agrawal et al. in~\cite{Agrawal:2018own}, but we fail to obtain the results of Heisenberg et al. in~\cite{Heisenberg:2018yae}, unless we use an incorrect procedure and place the center of the ellipses at $\Lambda$CDM instead of the center of the distribution obtained from the actual data.

Moreover, the authors of~\cite{Heisenberg:2018rdu} state: ``\dots giving the 3-$\sigma$ bound $\lambda\lesssim1.3$ required us (as it would require anybody else) to extrapolate from the 1- and 2-$\sigma$ contours. This can of course only be taken as an approximation to the inverse Fisher matrix. Doing so, we estimated an upper bound at the 3-$\sigma$ level, which should not be confused with a rigorous determination.'' Thus, they admit that their bound $\lambda < 1.35$ is not reliable. This problem does not exist in the consistent MCMC data analysis framework employed in the present work. In conclusion, we believe that extrapolating the existing $1\sigma$ and $2\sigma$ contours from current constraints to a $3\sigma$ one using the procedure of~\cite{Heisenberg:2018yae} is not correct, and one is instead required to perform a proper statistical analysis, by applying it either (ideally) directly to the quintessence models, as we have done here, or at least to a CPL parametrization of dark energy if one wants to follow a CPL-based, approximate procedure.

Concerning the future bounds on $\lambda$, let us stress here that the objective of the present paper has not been to perform a forecast analysis. Our main intention was to point out that the forcast of~\cite{Heisenberg:2018yae} has been restricted to the prospective constraints on the CPL parameters provided by the Euclid Red Book~\cite{Laureijs:2011gra}, while a combination of constraints from Euclid, Planck and the SKA is expected to provide stronger bounds; see~\cite{Sprenger:2018tdb}, particularly their Fig.~11. Moreover, the forecast analysis in~\cite{Sprenger:2018tdb} is not based on  the Fisher formalism, but on a Bayesian MCMC forecast technique, which is, in general, more accurate and more reliable. However, in order to rigorously estimate the future bounds on the cosmological models investigated here, one needs to perform either an MCMC-based forecast or a Fisher forecast directly on these models. This is in principle interesting, but is beyond the scope of the present paper.

Finally, it is also stated in~\cite{Heisenberg:2018rdu} that ``from the string theory point of view, the exact value of $\lambda$ is unknown, since only an order of magnitude $\mathcal{O}(1)$ can be given.'' We disagree with this statement, as we have shown in Section~\ref{models} that various string theory models predict specific values of $\lambda$, which allow one to rule out the models using existing cosmological observations, as done in this paper. Precise bounds on $\lambda$ can be reliably found, and we believe that knowing such bounds is relevant, at least for the purposes of the present paper. In conclusion, even though we agree that whether one uses the upper bound  $\lambda < 1$ or $\lambda < 1.35$, the conclusions in this paper do not change in application to the particular string theory quintessence models discussed in \cite{Obied:2018sgi,Agrawal:2018own}, we disagree with the authors of~\cite{Heisenberg:2018rdu} in that their $\mathcal{O}(1)$ estimates are sufficient. We believe that in the era of precision cosmology, small differences in constraints on models do matter, either for the existing theories or, sooner or later, for the theories to come; cf. Appendix~\ref{pres}.\footnote{In addition to the technical comments in~\cite{Heisenberg:2018rdu} on the present paper, which we believe we have addressed here, there is one additional comment stating that we have claimed in this paper that~\cite{Heisenberg:2018yae} has proposed ``string theory models of dark energy which are already ruled out by cosmological observations.'' We would like to state here that nowhere in the present paper has this claim been made. In fact we believe that none of the string theory quintessence models discussed in~\cite{Obied:2018sgi,Agrawal:2018own} were proposed by the authors of~\cite{Heisenberg:2018yae}.}


\section{The case of double-exponential potentials}\label{sec:doubleexp}

Here we observationally constrain models of quintessence with potentials of the form
\begin{equation}
V(\phi) = V_1 e^{\lambda_1 \phi} + V_2 e^{\lambda_2 \phi} \,.\label{eq:2exppot}
\end{equation}
This is important, as some of the string theory based models of quintessence discussed in Section~\ref{models} are of this form. Double-exponential quintessence models have already been studied in the literature; see, e.g.,~\cite{Barreiro:1999zs,Copeland:2006wr,Chiba:2012cb}. From the phenomenological point of view, the study of such potentials has been supported by the observation that they can exhibit a cosmologically viable {\it scaling} solution in an early epoch of the history of the universe, during which the energy density of the scalar field decreases proportionally to the energy density of matter or radiation, whichever is dominant.
\begin{figure}[t!]
\begin{center}
\includegraphics[scale=0.13]{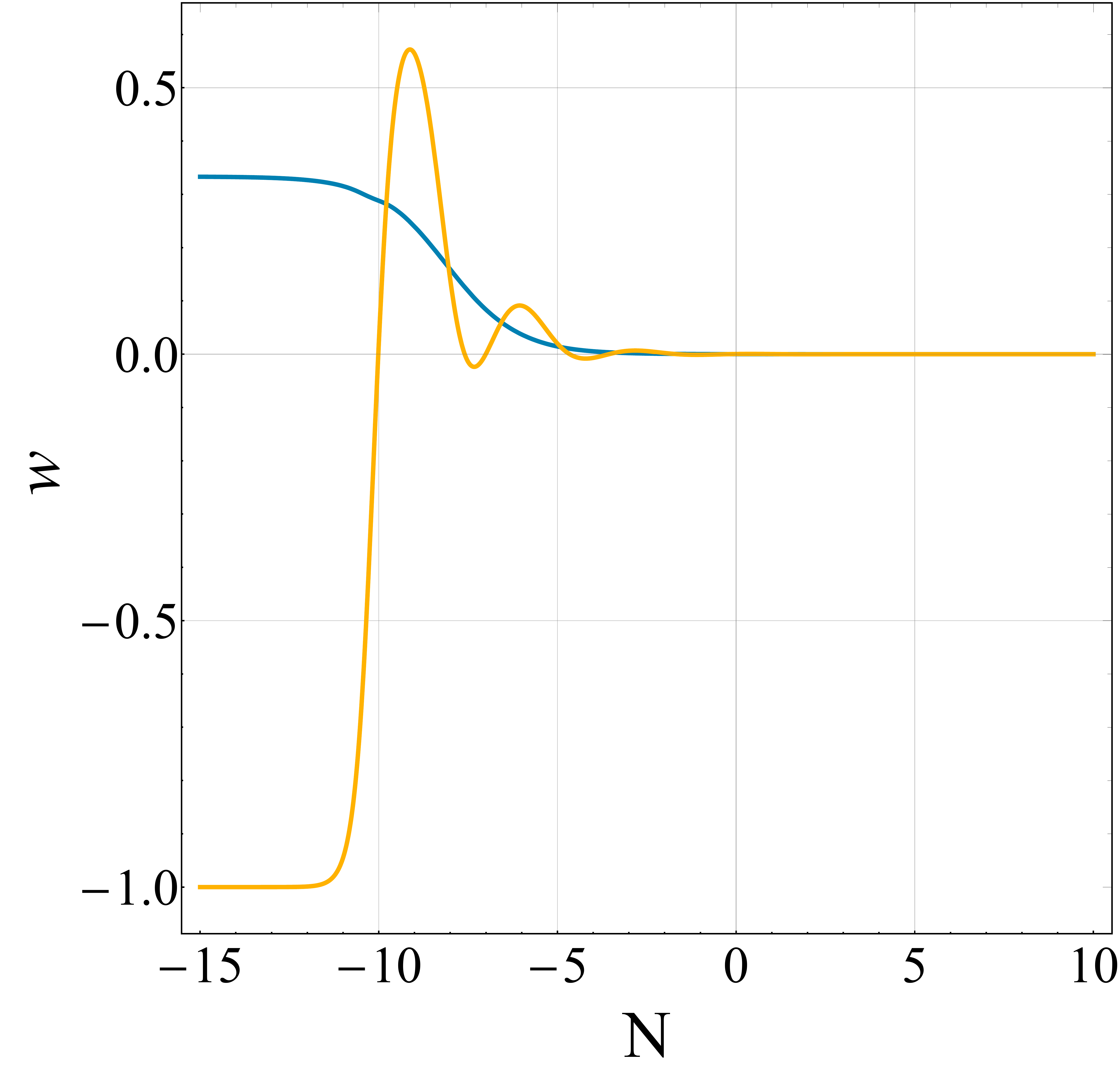}
\includegraphics[scale=0.13]{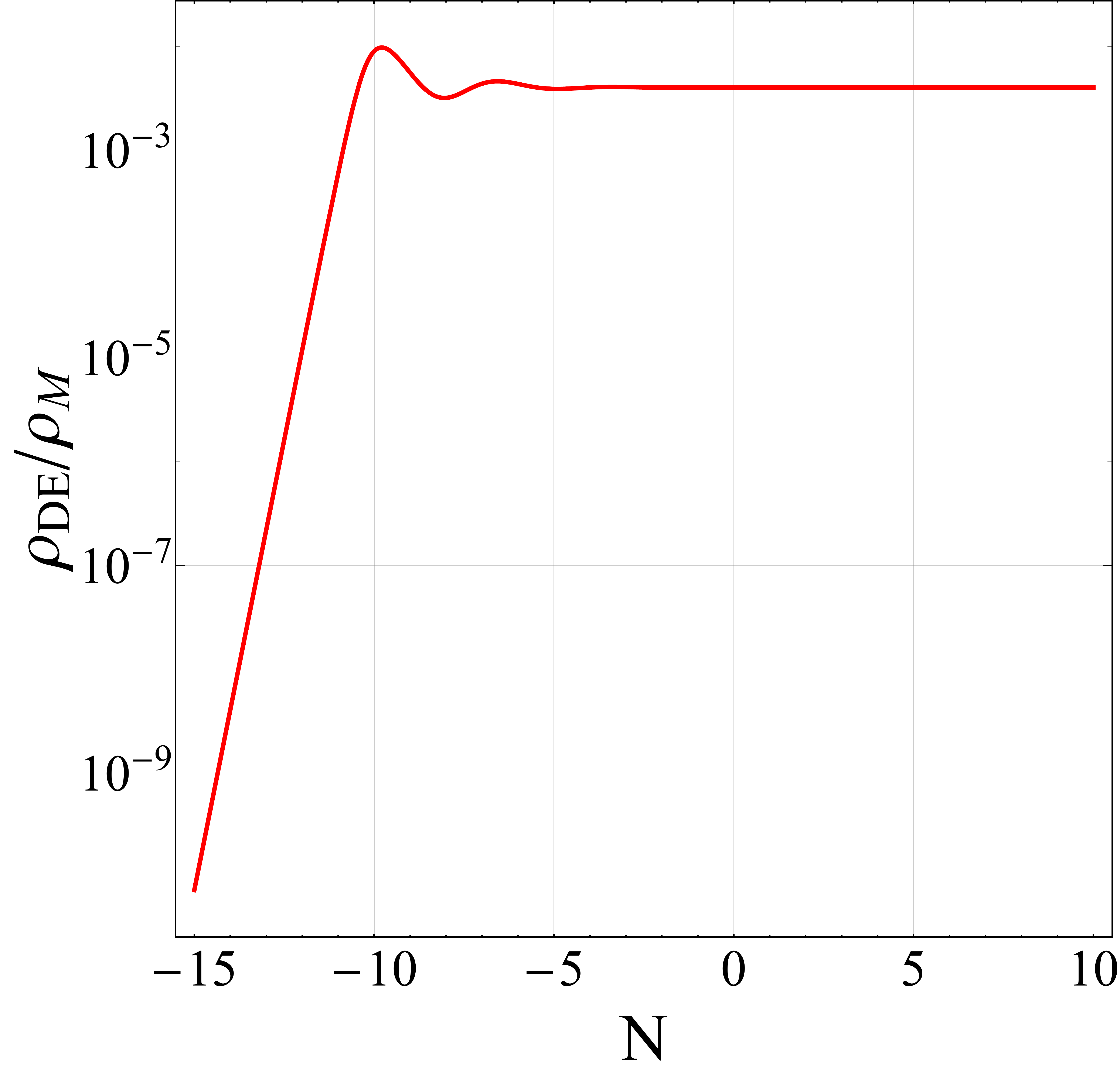}
\end{center}
\caption{\footnotesize An example of scaling solutions obtained through a single-exponential potential $V(\phi) = V_0 e^{\lambda \phi}$ with $\lambda^2 > 3(w_\text{B}+1)$. {\it Left panel:} The orange and blue curves show, respectively, the dark energy equation of state $w_\text{DE}$ and the effective (total) equation of state $w_\text{eff}$ as functions of $N\equiv\ln a$ ($N=0$ corresponds to today). {\it Right panel:} The ratio of the dark energy density $\rho_\text{DE}$ to that of matter $\rho_\text{M}$ as a function of $N$.}
\label{fig:single_exp_scaling}
\end{figure}

In fact, scaling fixed points exist already in the phase space of the single-exponential quintessence model. As shown in~\cite{Wands:1993zm,Copeland:1997et,PhysRevD.58.023503,Liddle:1998xm,Barreiro:1999zs,Copeland:2006wr}, for $\lambda^2 > 3(w_\text{B}+1)$, with $w_\text{B}$ being the equation of state for the background fluid in each epoch, the universe undergoes a scaling regime where the scalar field mimics the evolution of the barotropic fluid, with $w_\text{DE}=w_\text{B}$; the dark energy density parameter takes the form $\Omega_\text{DE}=3(w_\text{B}+1)/\lambda^2$. This scaling property is illustrated in Fig.~\ref{fig:single_exp_scaling} for a sufficiently large value of $\lambda$ (chosen to be $\sqrt{750}$ in this example). The left panel shows the equation of state for dark energy (in orange) and the effective equation of state for a universe dominated first by radiation (with $w_\text{B}=1/3$) and then by matter (with $w_\text{B}=0$). The figure shows that the scalar field, after some oscillations, quickly follows the background and one can achieve a scaling solution during matter domination in this example. By choosing a larger value of $\lambda$ one can push the beginning of this scaling period to earlier times, i.e. all the way to radiation domination. Such scaling solutions are interesting especially since they may provide a solution to the {\it coincidence problem}, i.e. why dark energy has an energy density close to that of matter at the present time; this is illustrated in the right panel of Fig.~\ref{fig:single_exp_scaling}, where we have shown the evolution of the quintessence energy density compared to that of matter. Even though these are very interesting features, the obvious problem, of course, is that a single-exponential potential has a constant slope, and therefore, once the scaling regime is switched on it never ends, hence there is no dark energy domination. This is consistent with the bound we found in Section~\ref{observations} on $\lambda$ through our comparison of the single-exponential model with the data. The overall $3\sigma$ bound of $\sim 1<\sqrt{3}$ on $\lambda$ shows that none of the scaling solutions survive the observational constraints.

\begin{figure}[t!]
\begin{center}
\includegraphics[scale=0.13]{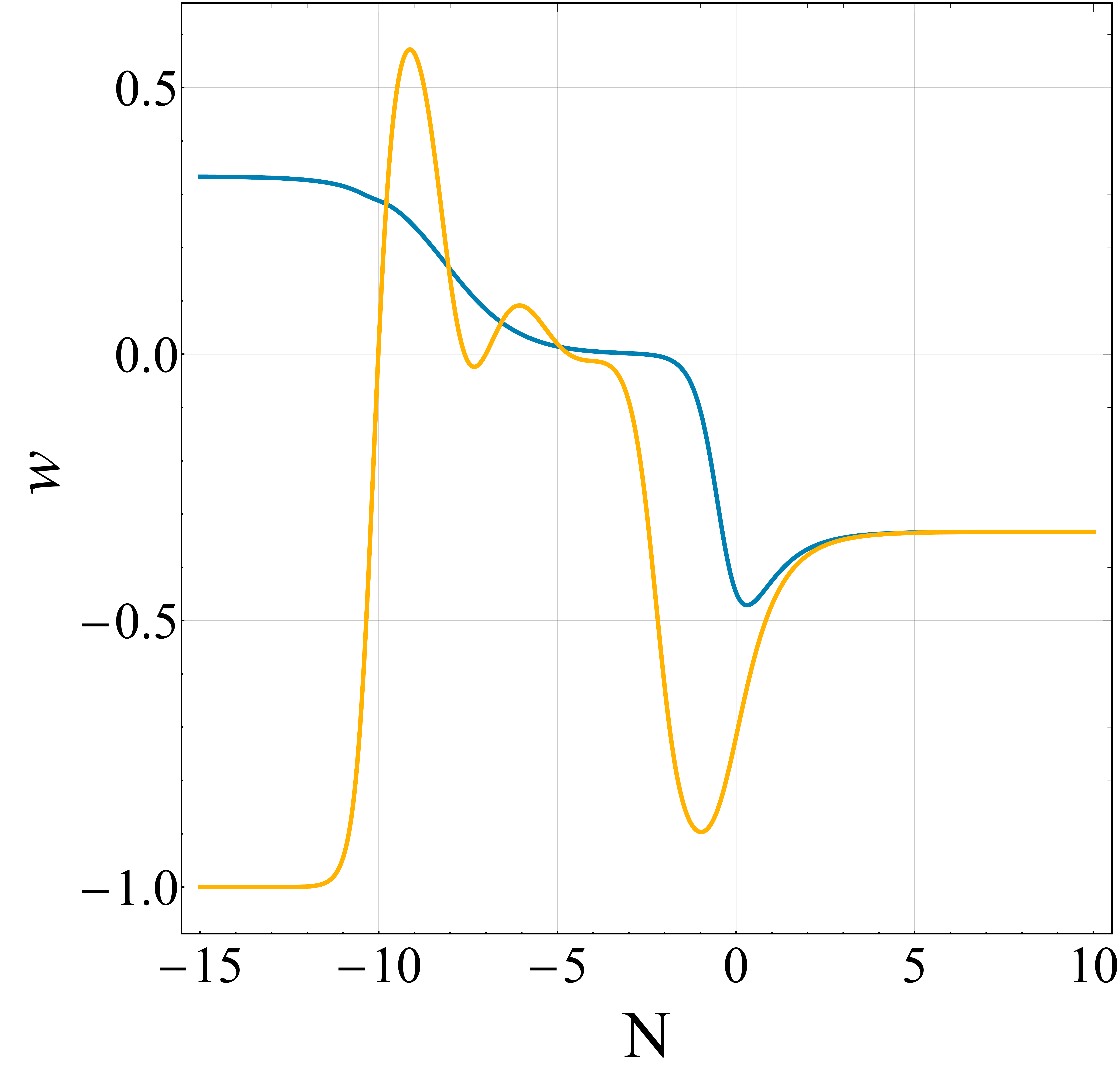}
\includegraphics[scale=0.13]{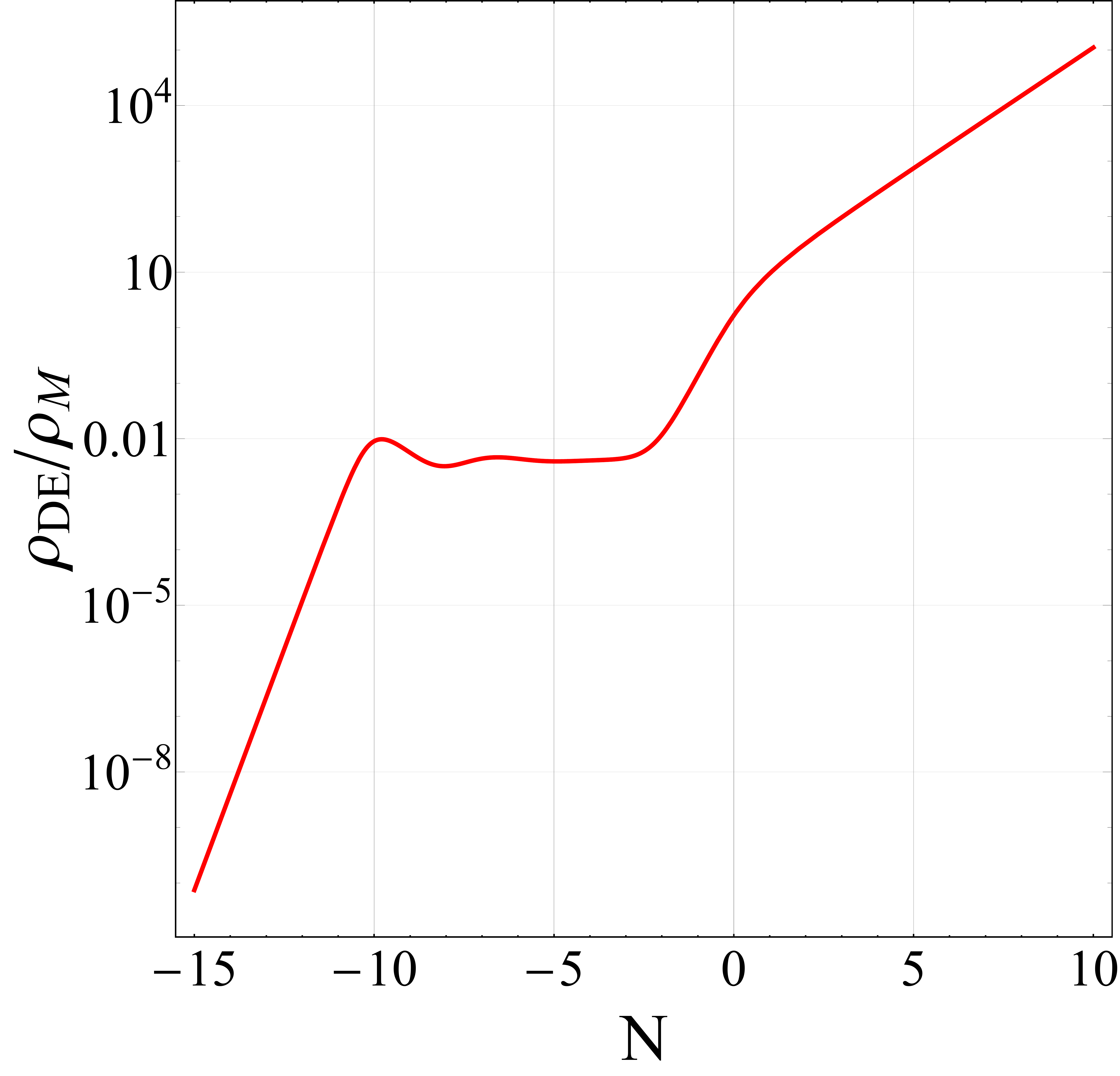}
\end{center}
\caption{\footnotesize The same as in Fig.~\ref{fig:single_exp_scaling} but for an example of scaling freezing solutions obtained through a double-exponential potential $V(\phi) = V_1 e^{\lambda_1 \phi} + V_2 e^{\lambda_2 \phi}$ with $\lambda_1^2 > 3(w_\text{B}+1)$ and $\lambda_2^2 < 3(w_\text{B}+1)$.}
\label{fig:double_exp_scaling}
\end{figure}

The phenomenological merit of the double exponentials is that under certain conditions the scaling solution can gracefully exit to the desired accelerating phase at late times, as demonstrated in~\cite{Barreiro:1999zs,Copeland:2006wr,Chiba:2012cb}. This transition can be obtained if $\lambda_1^2 > 3(w_\text{B}+1)$ and $\lambda_2^2 < 3(w_\text{B}+1)$ in the potential (\ref{eq:2exppot}).\footnote{Interestingly, the string theory examples with double-exponential potentials considered in Section~\ref{models} satisfy these conditions for both radiation and matter dominated epochs.} At early times, the potential is dominated by the $e^{\lambda_1 \phi}$ term, for which the scalar field follows the equation of state of radiation and/or matter, hence scaling solutions. Later in the evolution of the universe, the $e^{\lambda_2 \phi}$ term dominates, for which the evolution is not of the scaling form and the late-time attractor is the scalar field dominated solution (with $\Omega_\text{DE}=1$). In this scenario, the asymptotic value of the dark energy equation of state is $w_\text{DE}=-1+\lambda_2^2/3$, providing viable cosmologies, just as for the single exponential (\ref{eq:singexppot}) with $\lambda^2 < 3(w_\text{B}+1)$. Fig.~\ref{fig:double_exp_scaling} shows an example of this so-called {\it scaling freezing} scenario with the double-exponential potentials, where the transition from the scaling evolution to the scalar field dominated evolution has been depicted. The blue curve in the left panel of the figure shows how one can recover the three required epochs of radiation, matter and dark energy domination with a double-exponential potential, which was not possible through a single exponential with the scaling condition satisfied.

We now constrain the double-exponential quintessence models by performing an MCMC scan in the same way as we did in Section~\ref{observations} for the single-exponential case. The parameter space is now considerably more complex than the single-exponential model, as it contains different types of viable solutions depending on the values of $\lambda_1$ and $\lambda_2$. Since here we are not interested in particular classes of solutions and would only like to know the overall constraints on $\lambda_1$ and $\lambda_2$, we freely scan the parameter space of the model, i.e. over the parameters $V_0$ (where we have assumed $V_1=V_2\equiv V_0$), $\lambda_1$, $\lambda_2$ and $\phi_0$, as well as $\Omega_\text{M}$. As argued in~\cite{Barreiro:1999zs,Copeland:2006wr,Chiba:2012cb}, in principle one needs to also take into account constraints from early times, in particular the big bang nucleosynthesis bounds on $\Omega_\text{DE}$. This consideration requires $\lambda_1>5.5$~\cite{Barreiro:1999zs}, or even a tighter constraint of $\lambda_1>9.4$~\cite{Chiba:2012cb}, for the scaling freezing cases (assuming $\lambda_1>\lambda_2$). We however do not impose this {\it prior} on our parameters in the MCMC scans.

\begin{figure}[t!]
\begin{center}
\includegraphics[scale=0.6]{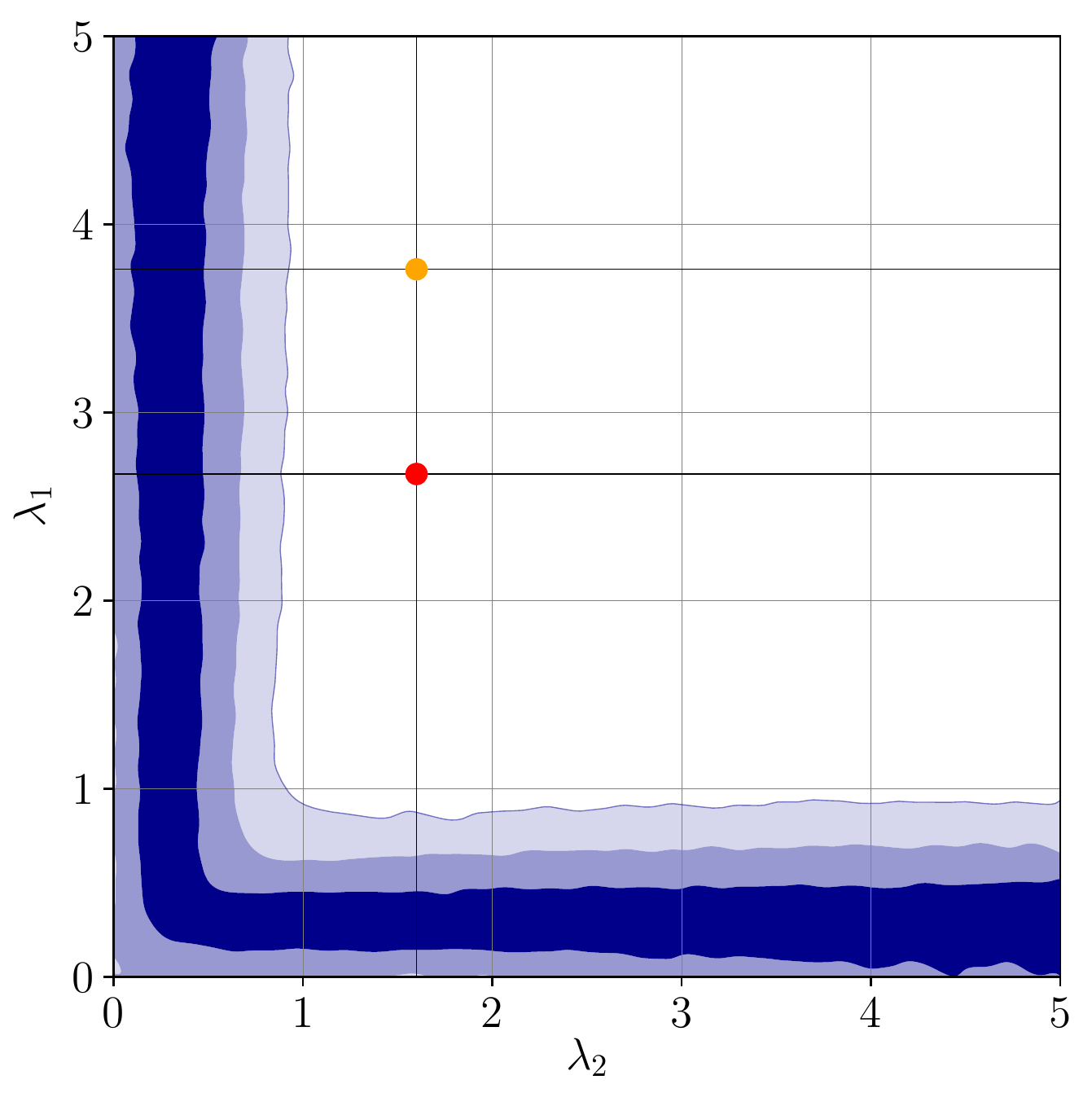}
\end{center}
\caption{\footnotesize Two-dimensional, marginalized constraints on $\lambda_1$ versus $\lambda_2$ for quintessence models with double-exponential potentials $V(\phi) = V_1 e^{\lambda_1 \phi} + V_2 e^{\lambda_2 \phi}$. The contours show $68\%$, $95\%$ and $99.7\%$ confidence levels. The two red and orange points on the plot correspond, respectively, to the two string theory based double-exponential models \rf{ob} and \rf{gutp} discussed in Section~\ref{mm}.}
\label{fig:double_exp_lambda1-2}
\end{figure}

Fig.~\ref{fig:double_exp_lambda1-2} shows the constraints in the $\lambda_1-\lambda_2$ plane as the result of our MCMC scan. As expected, the contours are symmetric in terms of the exchange of $\lambda_1$ and $\lambda_2$. Placing overall one-dimensional constraints on these two parameters is not useful, as depending on the value of one of them the constraint on the other changes. We can however recognize two general classes of viable solutions by looking at the two-dimensional contours of Fig.~\ref{fig:double_exp_lambda1-2}: 1) $\lambda_1^2 > 3(w_\text{B}+1)$ and $\lambda_2^2 < 3(w_\text{B}+1)$, or the opposite, where the universe starts with a scaling evolution and then transitions into a dark energy dominated epoch at late times, and 2) $\lambda_1^2, \lambda_2^2 < 3(w_\text{B}+1)$ where the universe never enters a scaling phase.

The two (red and orange) points in Fig.~\ref{fig:double_exp_lambda1-2} correspond to the two double-exponential models discussed in Section~\ref{mm}; we see again that these models are ruled out as their parameters are located far outside the $99.7\%$ confidence region. It is however important to note that these models can be ruled out even without comparing them with the full, statistical results of Fig.~\ref{fig:double_exp_lambda1-2}. The reason is that for both models one of the two $\lambda$ in the exponents is larger than $\sqrt{3}(w_\text{B}+1)^{1/2}$ and the other one is smaller, which means that the models are of the scaling freezing type (if both $V_{\mathcal{R}}$ and $V_{G}$ (or $\tilde V_{G}$) are nonzero). As we mentioned above, there are lower bounds on the larger $\lambda$ in these scaling freezing scenarios from big bang nucleosynthesis, which immediately rule out both models. One possibility of course is that one of the quantities $V_{\mathcal{R}}$ and $V_{G}$ (or $\tilde V_{G}$) vanishes. In this case, the models become of the single-exponential form, which are also ruled out since all the given values of $\lambda$ (i.e. $\sqrt{\frac{18}{7}}$, $\sqrt{\frac{50}{7}}$ and $\sqrt{14}$) are above the $3\sigma$ upper bound we found in Section~\ref{observations} for single exponentials. This is in line with our simple argument in Section~\ref{mm} for why these models are ruled out.

\section{More on the two-field model of Section~\ref{O16}}\label{sec:two-field}

In this appendix, we study the late-time dynamics of the two-field model of Section~\ref{O16}, given by the potential
\be\label{eq:two_field_pot}
V(\hat \rho, \hat \tau)= V_{\mathcal{R}} e^{- \sqrt{2\over 3} \hat \rho} e^{\sqrt{2} \hat \tau} +  V_{\Lambda} e^{\sqrt{6} \hat \rho} e^{ 2{ \sqrt {2}}\hat \tau} \ ,
\ee
for the canonical fields $\hat \rho$ and $\hat \tau$.

As explained in Section~\ref{O16}, the shallowest direction of this potential can effectively be described as a single-field exponential potential with a constant slope of $\lambda_\text{sh} \approx 1.6$. The late-time cosmological evolution along this direction is ruled out based on the results of Section~\ref{observations}. However, due to the field-space evolution from one steep direction to another, one might expect the appearance of some transient behavior which can potentially be viable. Here, we show that in practice this cannot be achieved, hence there are no viable cosmologies for this two-field model.

\begin{figure}[t!]
\begin{center}
\includegraphics[scale=0.15]{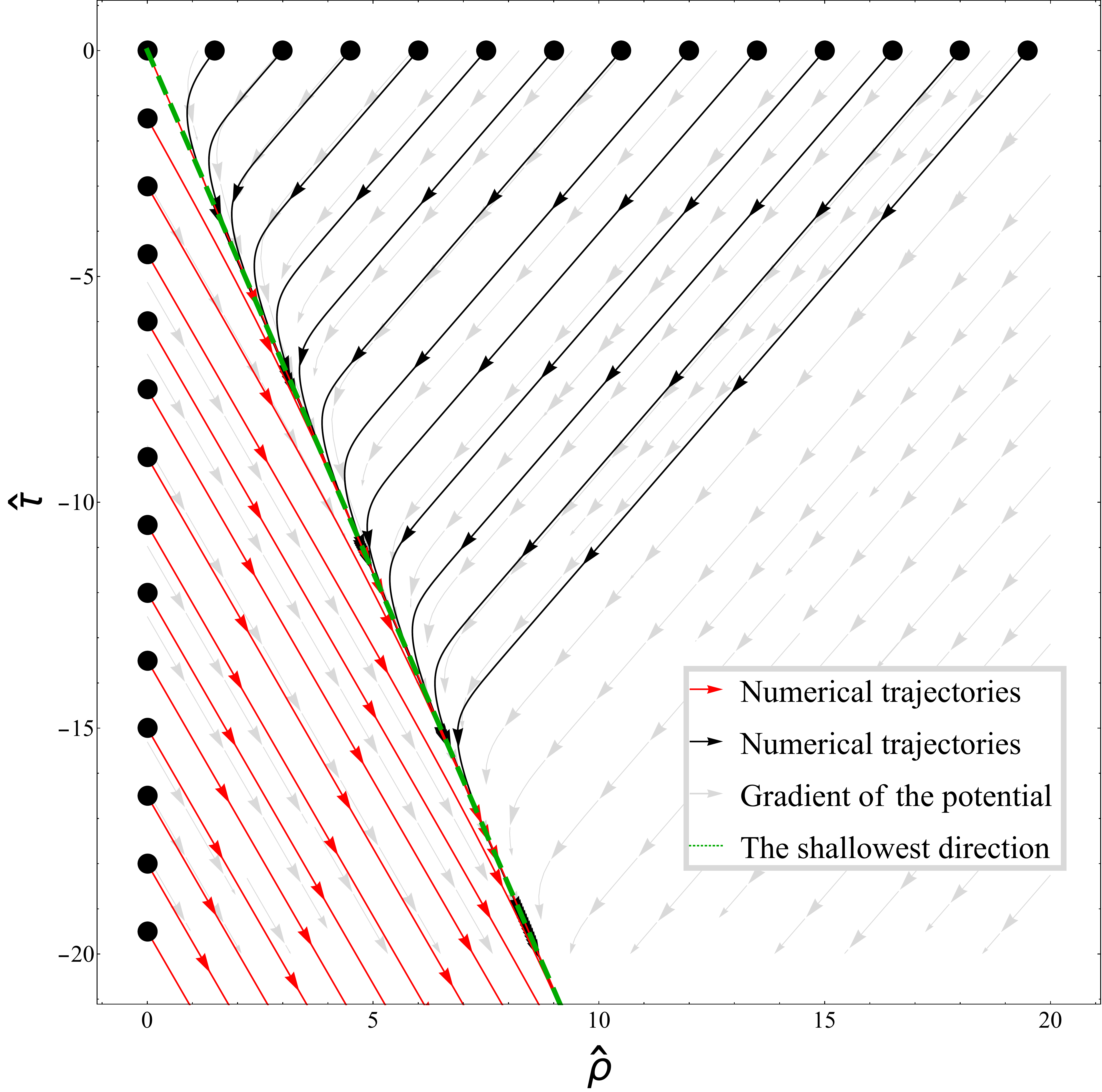}
\includegraphics[scale=0.16]{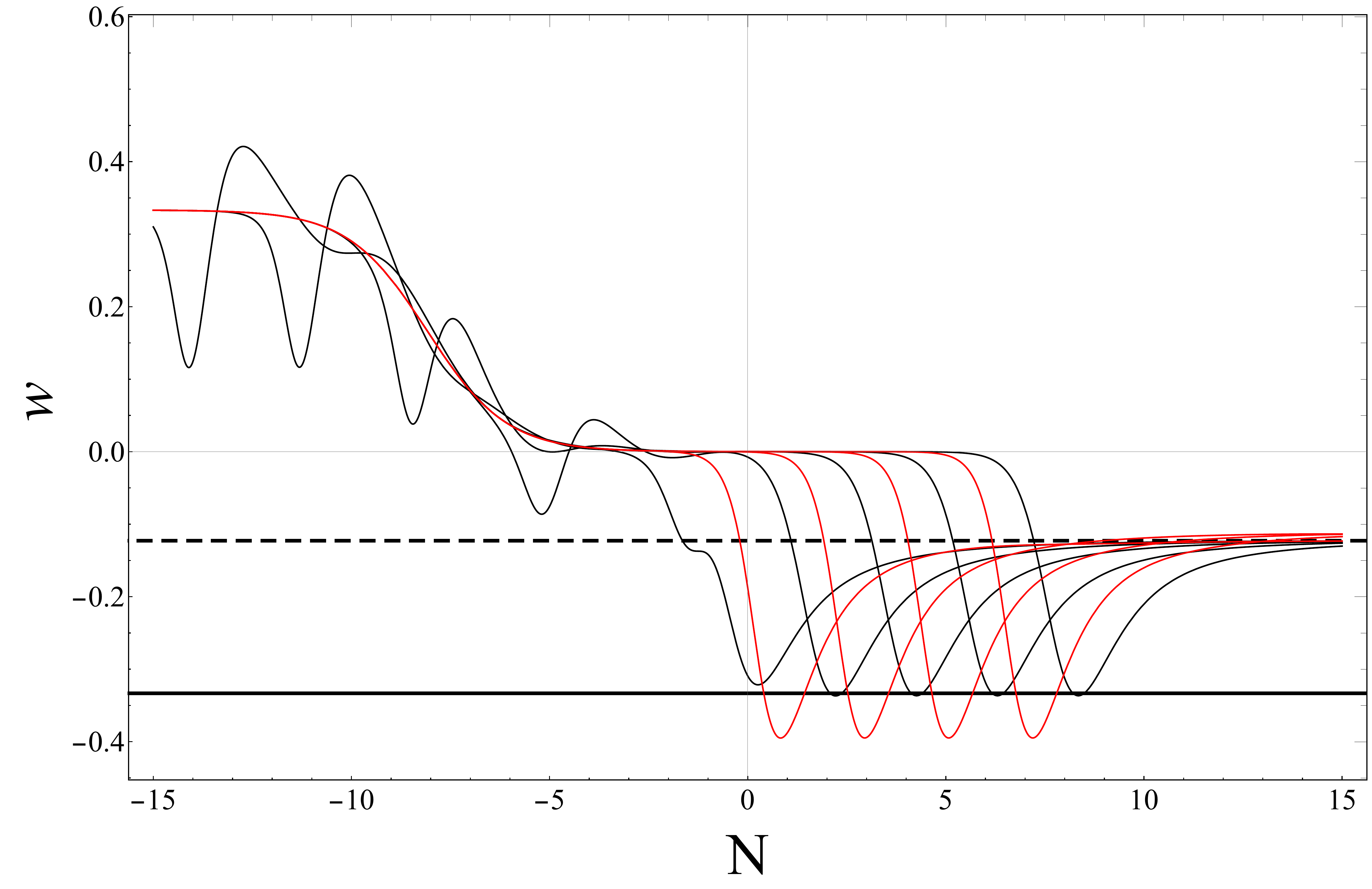}
\end{center}
\caption{\footnotesize In the left panel we show how the two-field system with the potential (\ref{eq:two_field_pot}) evolves in time. The difference with the left panel of Fig.~\ref{fig:2fields} is that here dark matter and radiation are present. The evolution is shown for $V_{\mathcal{R}} = 18 V_{\Lambda}$. In the right panel we show the evolution of the effective equation of state corresponding to the trajectories in the left panel (the black curves correspond to the black trajectories in the left plane, while the red ones correspond to the red trajectories). The horizontal dashed line corresponds to $-1 + \lambda^2_\text{sh}/3$.}
\label{fig:2fields-matter}
\end{figure}

In the left panel of Fig.~\ref{fig:2fields-matter}, we show the field trajectories in the field space, when the dark matter and radiation components are present. As in Fig.~\ref{fig:2fields}, here too the initial conditions for the fields are such that they are at rest initially.

In the right panel of Fig.~\ref{fig:2fields-matter}, we show the evolution of the effective equation of state corresponding to the trajectories in the left panel. In particular, the black curves in the right panel correspond to the black trajectories in the left plane, while the red ones correspond to the red trajectories. We have verified that the gradient of the potential along the red trajectories is sufficiently small to maintain thawing-type evolution, while most of the black trajectories (the ones starting far enough from the green dashed line) evolve in a very steep region above the green line, and therefore, resemble the scaling regime discussed in Appendix~\ref{sec:doubleexp}, which explains the wiggly behavior of the corresponding equation of state. From this figure it is clear that this potential cannot give cosmologically viable evolutions. Indeed, we see that both the red and black trajectories asymptotically reach the value corresponding to the shallowest direction (the horizontal dashed line), which in this case does not even give an accelerating universe (the horizontal solid line corresponds to $w_\text{eff} = -1/3$). We see, additionally, that there is a transient regime for all the trajectories, during which the equation of state is smaller than the asymptotic value given by $-1 + \lambda^2_\text{sh}/3$. This transient regime, in case of the red trajectories, gives accelerated expansion, which, however, is not rapid enough (the minimum of $w_\text{eff}$ reaches only values of $\sim -0.4$). For the case of black trajectories, the transient regime barely gives accelerated expansion, and therefore, those trajectories are also ruled out.

Even though the particular potential studied in this appendix is not viable, this two-field model exhibits \textit{phenomenologically} interesting features. Somewhat similar characteristics have been studied in the context of assisted inflation and assisted dark energy scenarios \cite{Liddle:1998jc,Tsujikawa:2006mw}. Similarly to the idea of double-exponential models of Section~\ref{sec:doubleexp}, where a scaling regime is achieved by the steep exponential followed by an acceleration regime when the shallower exponential dominates the dynamics, here the two-component field can roll along a steep direction and enter a scaling-type regime, and then continue rolling along a shallow direction, resulting in an accelerated universe.

\section{Evolution of precision in inflationary and dark energy parameters}\label{pres}

\begin{figure}[t!]
\begin{center}
\includegraphics[scale=0.7]{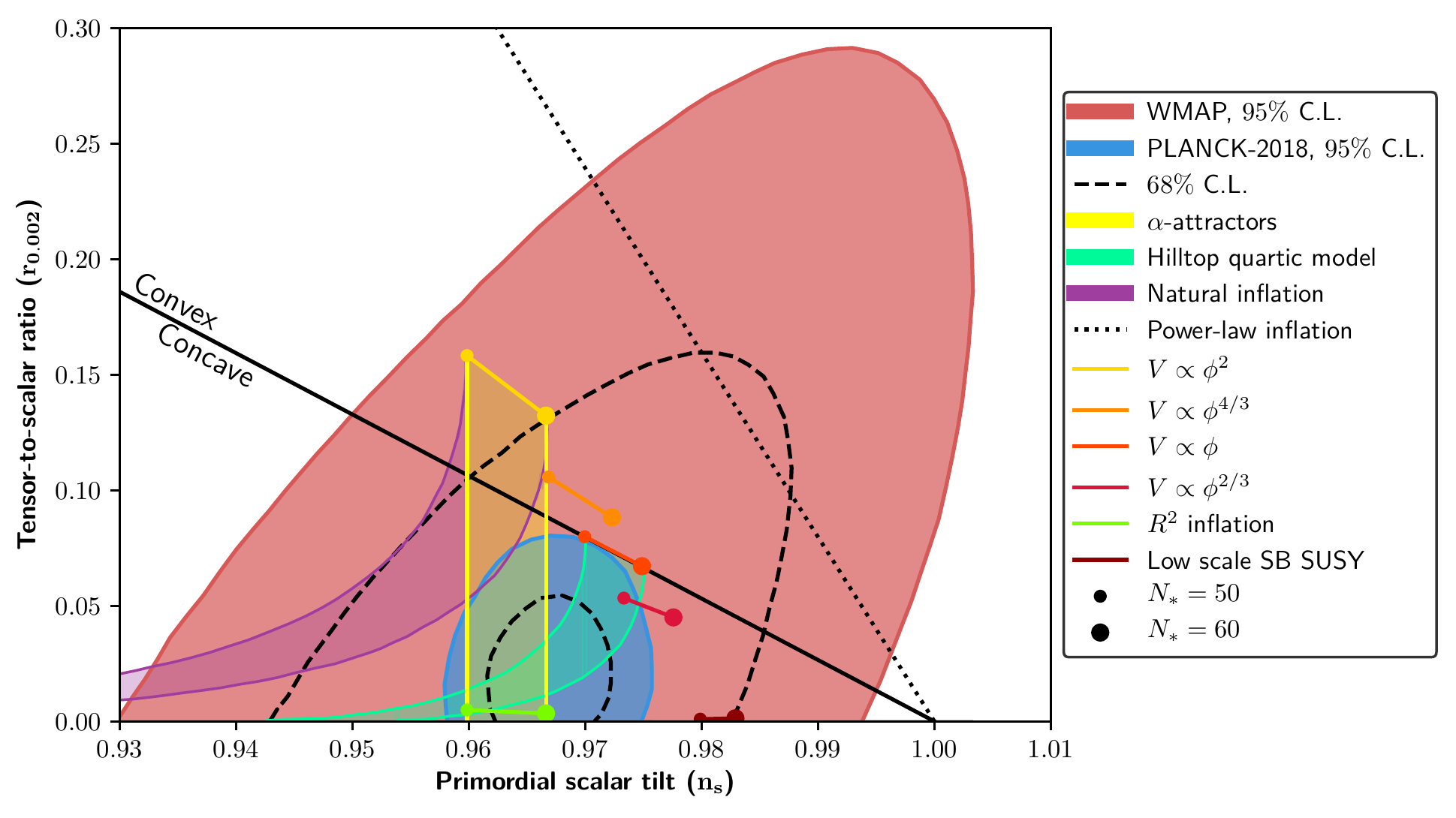}
\end{center}
\caption{\footnotesize Evolution of precision in inflationary parameters over a decade, from WMAP~\cite{Komatsu:2010fb} to Planck~\cite{Akrami:2018odb}. The reconstructed Planck constraints correspond to the combination TT,TE,EE+lensing+BK14+BAO provided in~\cite{Akrami:2018odb}. One can look, for example, at the area between $n_s=0.95$ and $n_s=0.98$. Although both of these values were inside the $68\%$ contour back in 2009, they are now strongly disfavored with more than $95\%$ confidence.}
\label{Precision}
\end{figure}

The current observational constraints on inflation by the CMB data, reconstructed in Fig.~\ref{Precision}, tell us that many favorite models of inflation, like the polynomial inflationary models of $\phi^k$ with $k=2, 1, 2/3$, are now disfavored. For polynomial models,
$n_s= 1 -{2+k\over 2N}$ and $ r={4k\over N}$, where $N$ is the number of $e$-foldings (between 50 and 60), while F-term and D-term models of inflation predict $n_s=0.98$, and racetrack inflation requires $n_s=0.95$. All these models were inside the $95\%$ sweet spot of the data in 2009 provided by the WMAP collaboration~\cite{Komatsu:2010fb}, as one can see in Fig.~\ref{Precision}. However, the figure shows that all the polynomial models, including $\phi^{2/3}$, are now either outside or close to the boundary of the $95\%$ confidence region of the Planck 2018 data. Similarly, the F-term and D-term inflationary models with $n_s=0.98$ or racetrack inflation with $n_s=0.95$ are now practically ruled out, and natural inflation is disfavored.

The improvement in the data came from two sources. The new Planck 2018 bound~\cite{Akrami:2018odb} on the primordial gravity waves, when combined with the BICEP2/Keck Array (BK14) constraints~\cite{Array:2015xqh}, is $r<0.064$ at the $2\sigma$ level. Another factor is the reduced error on $n_s$, which is now $0.004$, down from $0.006$ back in 2015. Many inflationary models which were consistent with the data in 2007-2009 are now either totally ruled out, or have become only marginally acceptable. This means that in the theoretical analysis of the data, it is no longer possible to perform a {\it parametric, order of magnitude analysis} as it was normal in the past, especially in string theory. The same concerns such expressions as `parametrically small', or `parametrically large'.
We can see examples in Fig.~\ref{Precision} showing that reducing the bound on $r$ from $\sim0.08$ to $\sim0.04$ has made various theoretical ideas either supported or ruled out by the precision data in cosmology.

Similarly, the constraints on the equation of state of dark energy become more and more precise. 15 years ago, the constraints on the parameter $c$, or $\lambda$ in the exponent of $e^{\pm \lambda \phi}$, allowed $c = \lambda = 1.6$ \cite{LopesFranca:2002ek,Kallosh:2002gf}.
Meanwhile, in the discussion of the quintessence models described in \cite{Obied:2018sgi} it was necessary to pay full attention to a small difference between numbers such as $c<1$ and $c <1.4$. Indeed, models with  $ 1 < c$ are ruled out by cosmological observations at least at the $3\sigma$ level, i.e. with more than $99.7\%$ confidence, whereas the condition  $c <1$ is not satisfied by the  string theory models of  \cite{Obied:2018sgi}.

\bibliographystyle{JHEP}

\bibliography{lindekalloshrefs,akramirefs}

\end{document}